\documentclass[aps,prd,reprint,showpacs,nofootinbib,floatfix,superscriptaddress]{revtex4-1}
\newcommand{\md}{V_{9+10}}
\newcommand{\feetwo}{f_{2}^{ee}}
\newcommand{\feethree}{f_{3}^{ee}}
\newcommand{\feesix}{f_{6+7}^{ee}}
\newcommand{\feeeight}{f_{8}^{ee}}
\newcommand{\feeeleven}{f_{11}^{ee}}
\newcommand{\feefourteen}{f_{14}^{ee}}
\newcommand{\feefifteen}{f_{15}^{ee}}
\newcommand{\feesixteen}{f_{16}^{ee}}
\newcommand{\fefour}{f_{\perp}^{ee}+f_{\perp}^{ep}+\left(\frac{A-Z}{Z}\right)f_{\perp}^{en}}
\newcommand{\fenine}{f_{r}^{ee}+f_{r}^{ep}+\left(\frac{A-Z}{Z}\right)f_{r}^{en}}
\newcommand{\fetwelve}{f_{v}^{ee}+f_{v}^{ep}+\left(\frac{A-Z}{Z}\right)f_{v}^{en}}
\newcommand{\fefourt}{f_{\perp}^{ee}+f_{\perp}^{ep}+f_{\perp}^{en}}
\newcommand{\feninet}{f_{r}^{ee}+f_{r}^{ep}+f_{r}^{en}}
\newcommand{\fetwelvet}{f_{v}^{ee}+f_{v}^{ep}+f_{v}^{en}}

\newcommand{\vall}{Eqs.~\ref{eq:v11}--\ref{eq:v1213}}
\newcommand{\sdots}{\hat{\sigma}_{1}\cdot\hat{\sigma}_{2}}
\newcommand{\scrosss}{\hat{\sigma}_{1}\times \hat{\sigma}_{2}}
\newcommand{\sdotr}{\hat{\sigma}\cdot\hat{r}}
\newcommand{\sdotv}{\hat{\sigma}\cdot\vec{v}}
\newcommand{\vcrossr}{\vec{v}\times\hat{r}}
\usepackage{graphicx}
\usepackage{bm}

\begin{document}

\title{Prospects for electron spin--dependent short--range force
  experiments with rare earth iron garnet test masses}
\author{T. M. Leslie}
\altaffiliation[Current address: ]{Department of Mathematics,
  Statistics, and Computer Science, University of Illinois at Chicago,
  Chicago, IL 60607}
\author{E. Weisman}
\affiliation{Department of Physics, Indiana University, Bloomington IN
  47405\\ and IU Center for Exploration of Energy and Matter,
  Bloomington IN 47408} 
\author{R. Khatiwada}
\affiliation{Department of Physics, Indiana University, Bloomington IN
  47405\\ and IU Center for Exploration of Energy and Matter,
  Bloomington IN 47408} 
\affiliation{Department of Physics, Indiana University-Purdue
  University Indianapolis, Indianapolis, Indiana 46202, USA}  
\author{J. C. Long}
\thanks{Corresponding author}
\affiliation{Department of Physics, Indiana University, Bloomington IN
  47405\\ and IU Center for Exploration of Energy and Matter,
  Bloomington IN 47408} 

\begin{abstract}
A study of the possible interactions between fermions assuming only
rotational invariance has revealed 15 forms for the potential
involving the fermion spins.  We review the experimental constraints
on unobserved macroscopic, spin--dependent interactions between
electrons in the range
below 1~cm.   An existing experiment, using 1~kHz mechanical oscillators as test
masses, has been used to constrain mass--coupled forces in this range.
With suitable modifications, including spin--polarized test masses,
this experiment can be used to explore 
all 15 possible spin--dependent interactions between electrons in this
range with unprecedented sensitivity.  Samples of ferrimagnetic
dysprosium iron garnet have been fabricated in the suitable test mass
geometry and shown to have spin densities on the order of
10$^{20}\hbar$/cm$^{3}$ with very low intrinsic magnetism.
\end{abstract}

\maketitle

\section{Introduction}
The possible existence of unobserved interactions of nature with
ranges from microns to millimeters and very weak
couplings to matter has begun to attract a great deal of scientific
attention.  Many theories beyond the Standard Model possess extended
symmetries that, when broken at high energy scales, lead to weakly
coupled, light bosons such as axions, familons, and Majorons, which
can generate relatively long--range interactions~\cite{ber12}.
Several theoretical attempts to explain dark matter and dark energy
also produce new weakly coupled long--range interactions.  The fact
that the dark energy density, of order (1~meV)$^{4}$, corresponds to a
length scale of $\sim$100~$\mu$m encourages searches for new
phenomena at this scale in particular~\cite{ade09}.  Particles which might
transmit such interactions are sometimes referred to generically as
WISPs (Weakly-Interacting Sub-eV Particles)~\cite{jae10} in recent
theoretical literature, or as ``portals'' to a hidden sector~\cite{pos08}.

A general classification of interactions
between non-relativistic fermions assuming only rotational invariance
reveals 16 different operator structures~\cite{dob06}.  Of these, 15 involve the
spin of at least one of the particles and 7 their relative momentum.
In general, experimental constraints on unobserved interactions
that depend on the spin and/or velocity of the particles are fewer and
less stringent than those for static, spin--independent
interactions~\cite{ade09}.  However, new
experimental results from initial searches for the former interactions
have accelerated over the last few years.  In particular, the
velocity--dependent interactions involving the spin of both particles
have been constrained at long range using the geomagnetic
field~\cite{hun13}, and at the atomic scale from an analysis of
spin--exchange interactions~\cite{jac10}. 

One approach to the search for short--range forces uses planar, 1~kHz
mechanical oscillators as test masses with a stiff conducting shield
in between them to suppress backgrounds~\cite{lon03}.  A
fully--constructed experiment in the lab of the authors uses tungsten
test masses to search for mass--coupled forces in the range below
1~mm.  With
modifications including spin--polarized test masses, this technique
can be used to create localized spin sources in close proximity with
non-zero relative velocity.  It thus has the capability to probe
essentially all of the spin and 
velocity--dependent interactions described in~\cite{dob06}, with
unprecedented sensitivity in the range of interest.  Ferrimagnetic rare
earth iron garnets show promise as spin--polarized test masses with
low intrinsic magnetism, and several samples have been fabricated in
the suitable geometry.

This paper is organized as follows. Sec.~\ref{sec:param} reviews the
parameterization in~\cite{dob06}, as applied to the proposed
spin--dependent force search.  The current short--range limits on
polarized electron interactions are reviewed in Sec.~\ref{sec:limits}.
The experiment, with details on polarized test masses made from
dysprosium iron garnet, is described in
Sec.~\ref{sec:expt}. Sensitivity calculations based on the available
test masses are presented in Sec.~\ref{sec:sens}.  

\section{\label{sec:param} Parameterization}
In the non--relativistic, zero--momentum transfer limit, the
long--range potential $V_{i}$ ($i$ = 1,...,16) in the general
classification in~\cite{dob06} for single boson exchange depends (in
the enumeration in~\cite{dob06}) on 72 dimensionless coupling
constants $f_{i}^{1,2}$.  Here, the superscripts denote the species of
interacting fermions. 

In the experiment described in Sec.~\ref{sec:expt}, the polarized
particles (that is,
the particles with non-zero projection of spin averaged over the
volumes of the test masses) are electrons.  There are nine components
of the spin--spin potential between two polarized electrons.  Three
are static, given (in SI units, and adopting the numbering scheme
in~\cite{dob06}) by:
\begin{widetext}
\begin{eqnarray}
V_{2} & = & \feetwo\frac{\hbar c}{4\pi}(\hat{\sigma}_{1}\cdot\hat{\sigma}_{2})\left(\frac{1}{r}\right)e^{-r/\lambda}
\nonumber \\
V_{3} & = & \feethree\frac{\hbar^{3}}{4\pi m_{e}^{2}c}
\left[(\hat{\sigma}_{1}\cdot\hat{\sigma}_{2})\left(\frac{1}{\lambda
      r^{2}}+\frac{1}{r^{3}}\right)-(\hat{\sigma}_{1}\cdot\hat{r})
      (\hat{\sigma}_{2}\cdot\hat{r}) 
      \left(\frac{1}{\lambda^{2} 
      r}+\frac{3}{\lambda r^{2}}+\frac{3}{r^{3}}\right)\right]e^{-r/\lambda}
\nonumber \\ 
V_{11} & = & -\feeeleven\frac{\hbar^{2}}{4\pi m_{e}}\left[(\hat{\sigma}_{1}\times\hat{\sigma}_{2})\cdot\hat{r}\right]\left(\frac{1}{\lambda r}+\frac{1}{r^{2}}\right)e^{-r/\lambda}.
\label{eq:v11}
\end{eqnarray}
\end{widetext}
Here, ${\vec{s}_{1,2}}=\hbar\hat{\sigma}_{1,2}/2$ are the spins of
electrons (in test masses 1 and 2), $\hat{r}=\vec{r}/r$ is the
unit vector along the direction between them, $\hbar$ is Planck's
constant, $c$ is the speed of light in vacuum, $m_{e}$ is the electron
mass, and $\lambda$ is the interaction range.
The remaining six components depend on the relative velocity $\vec{v}$ of the
electrons:
\begin{widetext}
\begin{eqnarray}
V_{6+7} & = & -\feesix\frac{\hbar^{2}}{4\pi m_{e}c}\left[(\hat{\sigma}_{1}\cdot\vec{v})(\hat{\sigma}_{2}\cdot\hat{r})\right]\left(\frac{1}{\lambda r}+\frac{1}{r^{2}}\right)e^{-r/\lambda}
\nonumber \\
V_{8} & = & \feeeight\frac{\hbar}{4\pi c}\left[(\hat{\sigma}_{1}\cdot\vec{v})(\hat{\sigma}_{2}\cdot\vec{v})\right]\left(\frac{1}{r}\right)e^{-r/\lambda}
\nonumber \\
V_{14} & = & \feefourteen\frac{\hbar}{4\pi}\left[\left(\hat{\sigma}_{1}\times\hat{\sigma}_{2}\right)\cdot\vec{v}\right]\left(\frac{1}{r}\right)e^{-r/\lambda}
\nonumber \\
V_{15} & = & -\feefifteen\frac{\hbar^{3}}{8\pi m_{e}^{2}c^{2}}\left\{\left[\hat{\sigma}_{1}\cdot(\vec{v}\times \hat{r})\right](\hat{\sigma}_{2}\cdot\hat{r})+(\hat{\sigma}_{1}\cdot\hat{r})\left[\hat{\sigma}_{2}\cdot(\vec{v}\times \hat{r})\right]\right\}\left(\frac{1}{\lambda^{2}r}+\frac{3}{\lambda r^{2}}+\frac{3}{r^{3}}\right)e^{-r/\lambda}
\nonumber \\
V_{16} & = & -\feesixteen\frac{\hbar^{2}}{8\pi m_{e}c^{2}}\left\{\left[\hat{\sigma}_{1}\cdot(\vec{v}\times \hat{r})\right](\hat{\sigma}_{2}\cdot\vec{v})+(\hat{\sigma}_{1}\cdot\vec{v})\left[\hat{\sigma}_{2}\cdot(\vec{v}\times \hat{r})\right]\right\}\left(\frac{1}{\lambda r}+\frac{1}{r^{2}}\right)e^{-r/\lambda}.
\label{eq:v16}
\end{eqnarray}
\end{widetext}

There are six components in the case where only one test mass is
polarized.  The potentials between a polarized electron and an
unpolarized atom of atomic number $Z$ and mass number $A$ are given
by:
\begin{widetext}
\begin{eqnarray}
V_{4+5} & = & -Z\left[\fefour\right]\frac{\hbar^{2}}{8\pi m_{e}c}\left[\hat{\sigma}_{1}\cdot(\vec{v}\times\hat{r})\right]\left(\frac{1}{\lambda r}+\frac{1}{r^{2}}\right)e^{-r/\lambda}
\nonumber \\
\md & = & Z\left[\fenine\right]\frac{\hbar^{2}}{8\pi m_{e}}(\hat{\sigma}_{1}\cdot\hat{r})\left(\frac{1}{\lambda r}+\frac{1}{r^{2}}\right)e^{-r/\lambda}
\nonumber \\
V_{12+13} & = & Z\left[\fetwelve\right]\frac{\hbar}{8\pi}(\hat{\sigma}_{1}\cdot\vec{v})\left(\frac{1}{r}\right)e^{-r/\lambda},
\label{eq:v1213}
\end{eqnarray}
\end{widetext}
where $\hat{r}$ points from the electron to the atom and $\vec{v}$
is their relative velocity. Following~\cite{dob06}, only one linear
combination of the
separate components in Eq.~\ref{eq:v1213} has been used (as in the
expression for $V_{6+7}$ in Eq.~\ref{eq:v16}), and the
coupling constants are given in terms of the $f_{i}^{1,2}$ by: 
\begin{eqnarray}
f_{\perp}^{1,2} & = & -f_{4}^{1,2}-f_{5}^{1,2}\nonumber \\
f_{r}^{1,2} & = & -f_{9}^{1,2}-f_{10}^{1,2}\nonumber \\
f_{v}^{1,2} & = & \; \; \: f_{12}^{1,2}+f_{13}^{1,2}.
\end{eqnarray}
The potentials $V_{11}$, $V_{12+13}$, and $V_{16}$
violate parity $(P)$, $V_{6+7}$ violates time--reversal
symmetry $(T)$, and $V_{9+10}$, $V_{14}$ and $V_{15}$ violate both $P$
and $T$. The potentials $V_{3}$ and $V_{9+10}$ are the dipole--dipole
and monopole--dipole interactions studied by Moody and
Wilczek~\cite{moo84}.  The remaining potential ($V_{1}$) corresponds to
the well-known Yukawa type between unpolarized objects, to which the
sensitivity of the experiment in Sec.~\ref{sec:expt} is discussed
elsewhere~\cite{lon03b}.

For the case of spin-0 or spin-1 boson exchange, the coefficients $f_{i}^{1,2}$
can be expressed in terms of the scalar and pseudoscalar couplings $g_{S},g_{P}$
or vector and axial couplings $g_{V},g_{A}$, respectively.  The case
of single massive spin-0 exchange is derived in~\cite{dob06}, as is
the case for spin-1 in the context of a massive $Z^{\prime}$ boson.
The results are summarized in Table~\ref{tab:coup},
with various simplifications, for the experiment in Sec.~\ref{sec:expt}.
\begin{table*}
\caption{\label{tab:coup} Coefficients $f_{i}^{1,2}$ in terms of
  scalar, pseudoscalar,
  vector, and axial coupling constants for the case of single massive spin-0
  and spin-1 boson exchange, following~\cite{dob06}, as applied to the
  experiment in Sec.~\ref{sec:expt}. The approximation
  $A = 2Z$ is used in Eq.~\ref{eq:v1213} for couplings to
  unpolarized masses, which for the case of the proposed experiment
  (which uses silicon masses) is accurate to within 
  1\%. The results for $\fefourt$ $(s = 1)$ and $\fetwelvet$ ignore
  additional terms scaled by $m_{e}/m_{p,n}$ and $m_{e}/M$, where $M$
  is explained in~\cite{dob06}.}
\begin{ruledtabular} 
\begin{tabular}{ccc}
\textbf{Parameter} & $\bm{s=0}$ & $\bm{s=1}$ \\ \hline 
$\feetwo$       &         0         & $(g_{A}^{e})^{2}$ \\            
$\feethree$     & $-\frac{1}{4}(g_{P}^{e})^{2}$ &
$\frac{1}{4}[(g_{V}^{e})^{2}+(g_{A}^{e})^{2}]$ \\
$\feeeleven$    &         0         & $g_{A}^{e}g_{V}^{e}$ \\            
$\feesix$       &         0         & $g_{A}^{e}g_{V}^{e}$\footnotemark[1] \\
$\feeeight$     &         0         & $-\frac{5}{4}(g_{A}^{e})^{2}$ \\
$\feefourteen$  &         0         & $(g_{A}^{e})^{2}$\footnotemark[1] \\
$\feefifteen$   &         0         & $(g_{V}^{e})^{2}$\footnotemark[1] \\
$\feesixteen$   &         0         & $g_{A}^{e}g_{V}^{e}$\footnotemark[1] \\
$\fefourt$      &
$\frac{1}{2}g_{S}^{e}[g_{S}^{e}+g_{S}^{p}+g_{S}^{n}]$ & $\frac{1}{2}[3(g_{V}^{e})^{2}+(g_{A}^{e})^{2}+g_{V}^{e}g_{V}^{p}+g_{V}^{e}g_{V}^{n}]$ \\
$\feninet$      & $g_{P}^{e}[g_{S}^{e}+g_{S}^{p}+g_{S}^{n}]$ & $g_{A}^{e}[g_{V}^{e}+g_{V}^{p}+g_{V}^{n}]$\footnotemark[1] \\
$\fetwelvet$    &         0         &
$2g_{A}^{e}[2g_{V}^{e}+g_{V}^{p}+g_{V}^{n}]$ \\
\end{tabular}
\end{ruledtabular}
\footnotetext[1]{This is the more generic notation used or implied in~\cite{hun13}.}   
\end{table*}

\section{\label{sec:limits} Experimental limits}
\begin{figure}
\includegraphics[width=3.3in]{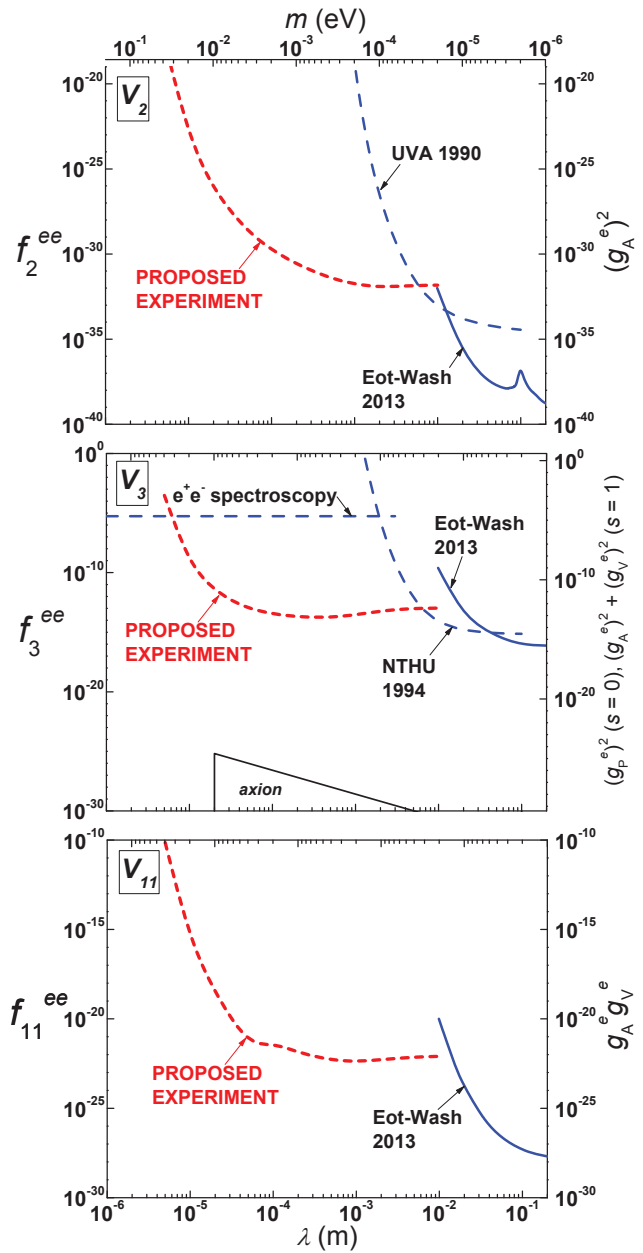}
  \caption{\label{fig:ppstatic} Projected sensitivity of proposed
    experiment to static
    spin--spin interactions (Eq.~\ref{eq:v11}), with current
    limits and theoretical prediction. Interaction
    strength according to all parameterizations in
    Table~\ref{tab:coup} is plotted versus the range
    $\lambda$ (lower axes) and the mass of an unobserved
    boson (upper axes). Excluded regions are above
    the curves. For $V_{2}$,
    solid curve is the 1~$\sigma$ direct
    limit on $(g_{A}^{e})^{2}$~\cite{hec13}, also expressed as
    $\feetwo$.  Dashed curve is the limit from~\cite{rit90}
    re-interpreted in terms of Eq.~\ref{eq:v11}. For $V_{3}$,
    bold solid curve is
    the direct limit on $(g_{P}^{e})^{2}$~\cite{hec13},
    also re-scaled to $\feethree$ in Eq.~\ref{eq:v11}. Dashed curves are
    the limit from~\cite{ni94} and the anomaly
    from~Refs.~\cite{mil75}--\cite{hil01}
    re-interpreted in terms of Eq.~\ref{eq:v11}; thin solid
    curve is the prediction for the axion~\cite{moo84}. For $V_{11}$,  
    solid curve is the direct
    limit on $g_{A}^{e}g_{V}^{e}$~\cite{hec13}, also expressed as
    $\feeeleven$.}
\end{figure}

Fig.~\ref{fig:ppstatic} shows the
experimental limits on static spin--spin interactions between
electrons (Eq.~\ref{eq:v11}) in the range between 1~$\mu$m and 10~cm. The best
limits above 1~cm derive from the spin--polarized torsion
pendulum experiment in the Eot--Wash group at the University of
Washington, previously used to constrain spin--dependent forces at
terrestrial and astronomical ranges~\cite{hec08}.  The ``spin
pendulum'' consists of an array of Alnico and SmCo$_{5}$ permanent
magnets arranged so that the orbital moments in the latter cancel the
spin moments in the former, resulting in a polarized test 
mass with negligible external field.  A recent shorter--range version
of this experiment~\cite{hec13} used a set of similarly--designed spin sources
placed 15-20~cm from the pendulum, arranged in
several configurations to enhance sensitivity to $V_{2}$, $V_{3}$, and
$V_{11}$ in Eq.~\ref{eq:v11}.  The results appear to be the
first short--range limits for electrons interpreted directly in terms
of these potentials.  They are reported in~\cite{hec13} as limits on
the couplings $(g_{A}^{e})^{2}$, $(g_{P}^{e})^{2}$, and
$g_{A}^{e}g_{V}^{e}$, respectively, and are shown in
Fig.~\ref{fig:ppstatic} according to those parameterizations and the
$f_{i}^{ee}$. The limits 
on $\feetwo$ and $\feethree$ are 1-4 orders of magnitude more
sensitive than previous results in the range near 1~cm, and the limit
on $\feeeleven$ appears to be the first such constraint in the range
of interest. 

Fig.~\ref{fig:ppstatic} also shows the limits on $\feetwo$ that can be
derived from the spin--polarized torsion pendulum at the University of
Virginia~\cite{rit90}.  The spin sources in this experiment consisted
of compensated rare earth ferrimagnets, which inspired the proposed
experiment in Sec.~\ref{sec:expt}, in the form of powder pressed into
high--permeability cylinders and polarized along their symmetry axes.
The results of the
original experiment are reported in terms of a fraction $\alpha$
of the strength of the (infinite--ranged) magnetic dipole--dipole
interaction between electrons:
\begin{equation}
\alpha=(1.6 \pm 6.9)\times 10^{-12}.
\label{eq:uva}
\end{equation}
The test cylinders were oriented side-by-side
with their axes parallel, a configuration which strongly suppressed
the $\hat{\sigma}\cdot\hat{r}$ terms in the dipole--dipole potential
and in which the finite--sized test masses could be approximated by
point dipoles up to correction factors of order unity.  The curve in
Fig.~\ref{fig:ppstatic} is thus obtained by converting the limit on
$\alpha$ to a magnetic dipole--dipole energy and equating it to
the expression for $V_{2}$ in Eq.~\ref{eq:v11}, where $r$ is fixed at
the 3.4~cm test mass separation reported in~\cite{rit90}.  The
long--range limit of the curve corresponds to the result reported for
this experiment in~\cite{dob06}.\footnote{The exact long-range limit
  is stronger than the result in~\cite{dob06},
  on account of an apparent error in Eq.~4.12 of that reference, at
  least partially confirmed by the authors.  The term containing the
  fine structure constant in that equation is mis-scaled by a factor of
  $4\pi$~\cite{dob13}.  This is compensated somewhat by the larger
  value of $r$ (10~cm) assumed in~\cite{dob06} for the experiment
  in~\cite{rit90}.}  

Similarly, short--range limits on $\feethree$ can be derived from the
experiment by Ni and co-workers at the National Tsing Hua University in
Taiwan~\cite{ni93,chu93,ni94}.  This experiment used a SQUID
magnetometer to monitor 
the interaction between spin--polarized test masses (also consisting of
compensated rare earth ferrimagnets) and a sample of paramagnetic
salt, as the test masses were rotated around the sample at a distance
of about 5~cm.  The results of this experiment are also reported in
terms of the electron magnetic dipole--dipole
interaction.\footnote{As noted in~\cite{dob06}, the explicit
  potential, which appears in~\cite{ni93} and~\cite{chu93}, scales as $1/r$, as
opposed to the expected $1/r^{3}$.  The authors of~\cite{dob06} suspect this
to be a typographical error, which has been confirmed~\cite{ni13}.}  The most
sensitive result~\cite{ni94} is:
$\alpha_{s}=(1.2 \pm 2.0)\times 10^{-14}$. The test mass polarization
was oriented either directly toward or away from the salt, maximizing
the contribution from the $\hat{\sigma}\cdot\hat{r}$ terms. The curve in
Fig.~\ref{fig:ppstatic} is thus obtained by converting $\alpha_{s}$ to a
magnetic dipole--dipole energy and equating it to the expression for
$V_{3}$ in Eq.~\ref{eq:v11}, with $r$ fixed at 5~cm.  Again, the
long--range limit of the curve corresponds to the result reported for
this experiment in~\cite{dob06}.\footnote{The corresponding result
  in~\cite{dob06}, Eq.~4.10, contains the same order-of-magnitude ($4\pi$)
  error as Eq.~4.12.  The error is also present in Eq.~4.11. The
  limits on the vector, axial, and pseudoscalar couplings derived from
  these results (Eqs. 5.32, 5.34 and 6.4 of~\cite{dob06}) should be
  scaled accordingly.} 

Below about about 2~mm, stronger limits on $\feethree$ can be inferred
from precision measurements of the hyperfine splitting in the ground state
of positronium~\cite{mil75,mil83,rit84}.  
There is currently a $\sim 4\sigma$ difference between these
measurements and QED theory~\cite{kni00,mel01,hil01}. The
horizontal line in the middle plot in Fig.~\ref{fig:ppstatic} results
from equating the energy discrepancy to the expression for $V_{3}$ in
Eq.~\ref{eq:v11}, with $r$ fixed at the positronium Bohr radius (0.1~nm). 
An analogous analysis of the same system has been used to constrain
unparticles~\cite{lia07}.

\begin{figure}
\includegraphics[width=3.3in]{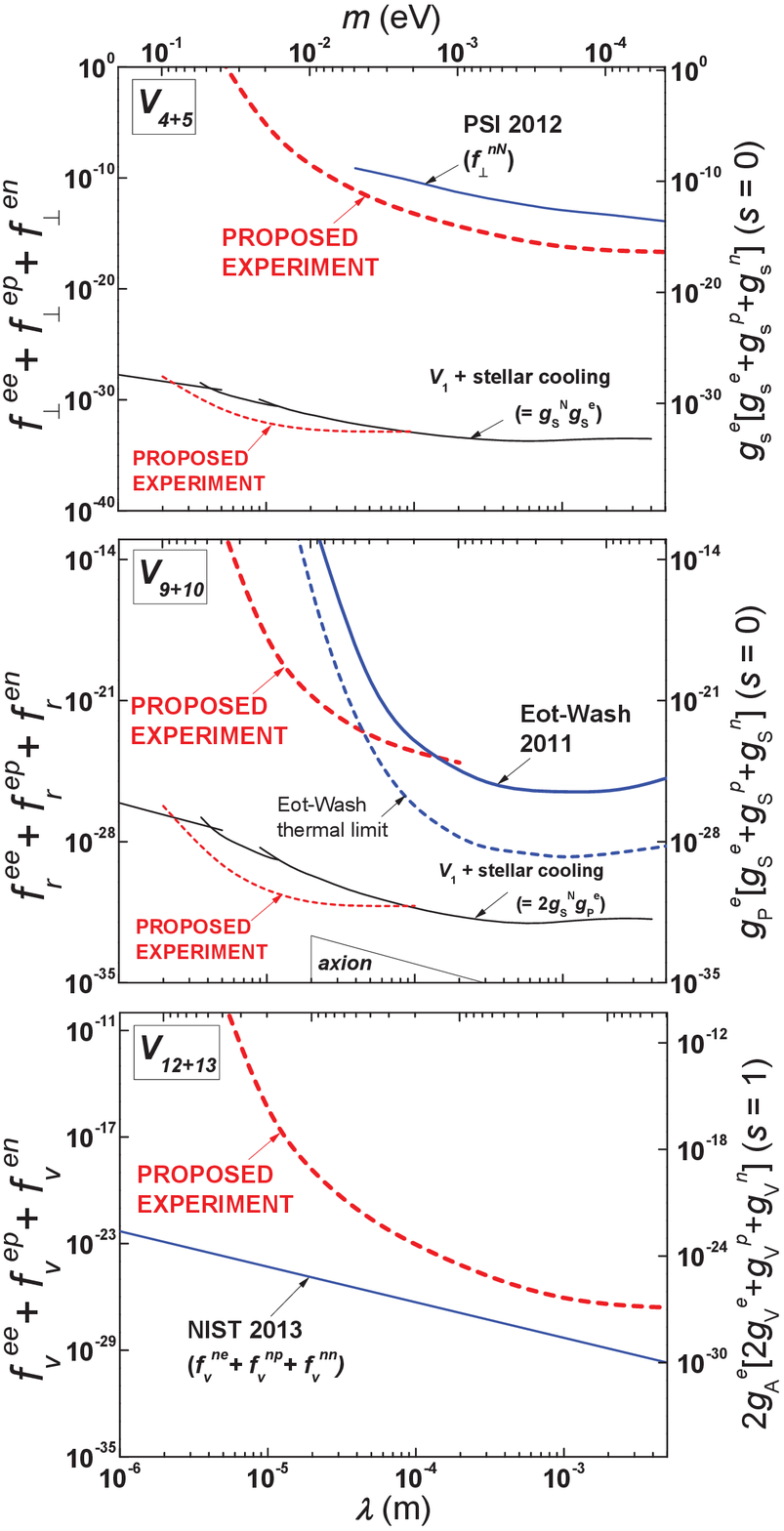}
  \caption{\label{fig:up} Projected sensitivity of proposed experiment to
    interactions between polarized and unpolarized
    particles (Eq.~\ref{eq:v1213}), with current limits and
    theoretical prediction.  For comparison, solid curves in the
    $V_{4+5}$ and $V_{12+13}$ plots are the direct limits
    (2~$\sigma$ and 1~$\sigma$, respectively)
    for the case of polarized neutrons~\cite{pie12,yan13}. For
    $V_{9+10}$, the bold
    solid curve is the 2~$\sigma$ direct limit on
    $g_{S}^{N}g_{P}^{e}$~\cite{hoe11}, also rescaled to $\feninet$ in
    Eq.~\ref{eq:v1213}; bold dashed curve is the projected thermal
    limit. Thin solid curve is the prediction for the
    axion~\cite{moo84}. Lower solid curves in the $V_{4+5}$ and
    $V_{9+10}$ plots are the indirect limits inferred from stellar
    cooling arguments~\cite{raf12}; the additional projected curves
    show expected improvements from the proposed experiment with
    unpolarized test masses~\cite{lon03b}.}  
\end{figure}

The $V_{3}$ plot also shows the prediction for the axion (for the
case of a spin-0 interaction), for which there exists an explicit
relationship between the coupling strength and the range.  The value
is derived from~\cite{moo84}, and also re-scaled to $\feethree$ according to
Table~\ref{tab:coup}.  The cutoff at 10~meV is the limit inferred from
SN1987a~\cite{ros00}.  As noted in Ref.~\cite{vis14}, the remaining
axion prediction in Fig.~\ref{fig:ppstatic} (and Fig.~\ref{fig:up}) is
allowed even if the recent BICEP2 measurement of the tensor-to-scalar
ratio in the cosmic microwave background~\cite{bic14} is correct,
lending additional interest to this part of the parameter space. 

\begin{figure*}
\includegraphics[width=6.3in]{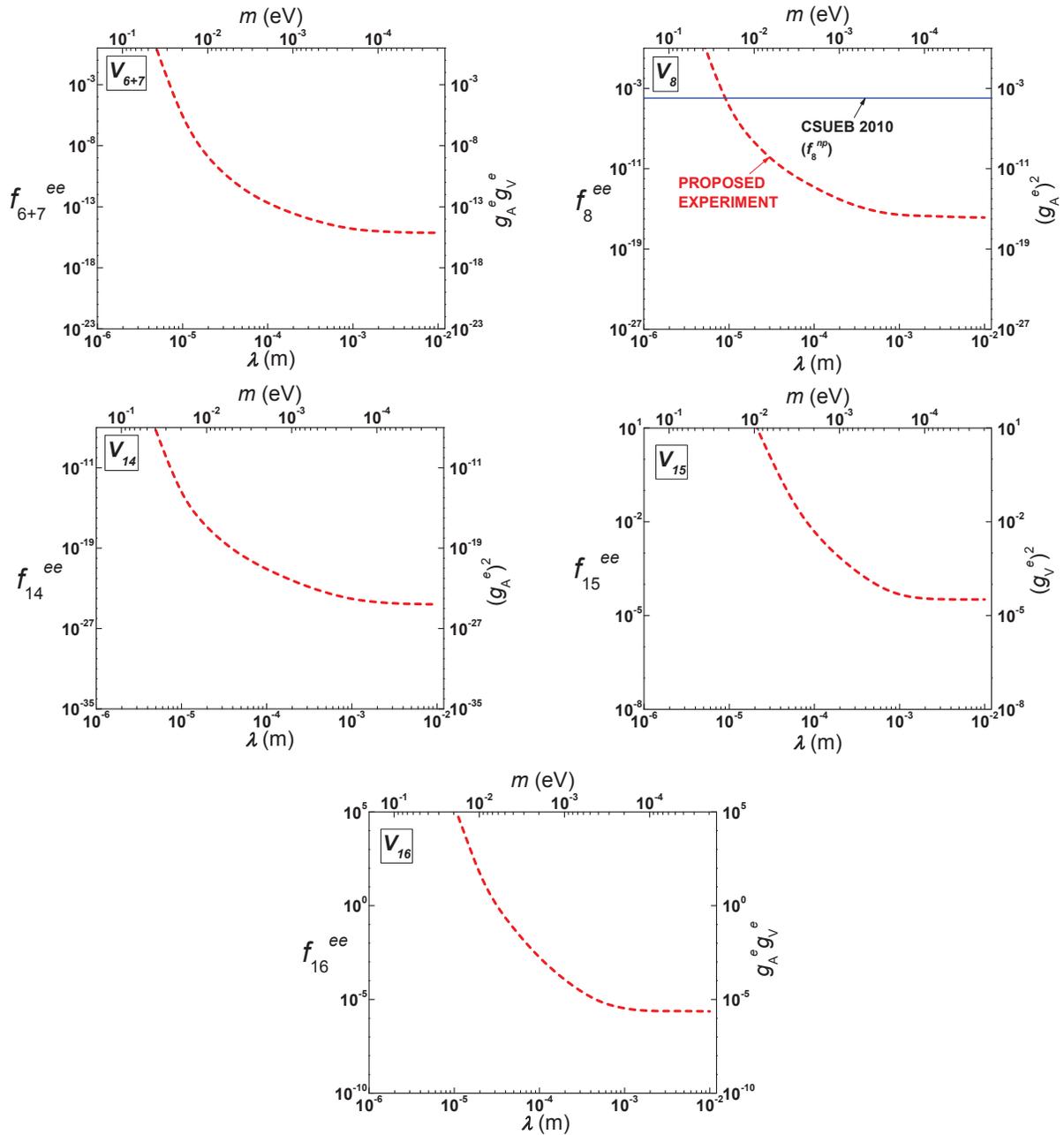}
  \caption{\label{fig:ppdyn} Projected sensitivity of proposed experiment to
    velocity--dependent 
    spin--spin interactions (Eq.~\ref{eq:v16}). For comparison, the
    solid curve in the $V_{8}$ plot is an extension of the 2~$\sigma$
    limit on $g_{V}^{n}g_{V}^{p}$ for nucleons~\cite{jac10}, also
    re-scaled to $f_{8}^{np}$ in Eq.~\ref{eq:v16}.}
\end{figure*}

Fig.~\ref{fig:ppdyn} shows the limits on velocity--dependent spin--spin
interactions (Eq.~\ref{eq:v16}) in the range of interest.  For the case
of electrons, these interactions appear to be unconstrained in this
range.  At $\lambda=1$~km, the lower limit of the range analyzed
in~\cite{hun13}, the constraints on electron interactions range from
$10^{-32}$--$10^{-22}$ for the case of $V_{14}$ and $V_{8}$, to
$10^{-7}$--$10^{-1}$ for the case of $V_{16}$ and $V_{15}$, with the
remaining interactions constrained at $10^{-17}$--$10^{-12}$. 

For comparison, the solid line in the $V_{8}$ plot is the limit calculated for
the nucleon coupling $f_{8}^{np}$ by a California State
University-East Bay collaboration, based on the analysis of
atomic spin exchange interaction cross sections~\cite{jac10}.  The
analysis compared the theoretical cross sections, calculated with the usual
spin--dependent electromagnetic potentials responsible for
spin exchange replaced with potentials of the form in
Eqs.~\ref{eq:v11} and~\ref{eq:v16}, with
data from He--Na collisions.  The result for $V_{8}$ is reported in~\cite{jac10}
as a limit on the coupling $g_{A}^{n}g_{A}^{p}$, and has been re-scaled in
Fig.~\ref{fig:ppdyn} according to Table~\ref{tab:coup}, with the
additional substitution ${\vec{s}_{2}}=3\hbar\hat{\sigma}_{2}/2$ in
the equation for $V_{8}$ to account for the Na nuclei which
carried the proton spin.  The limit has also been extended beyond the
micron range reported in~\cite{jac10}.

Short--range limits on the interactions in Eq.~\ref{eq:v1213}
are shown in Fig.~\ref{fig:up}. The velocity--dependent interactions
$V_{4+5}$ and $V_{12+13}$ appear to be unconstrained for the case of
polarized electrons. For comparison, the solid line in the $V_{4+5}$
plot is the limit on the corresponding coupling for polarized nucleons from an
experiment at the Paul Scherrer Institute~\cite{pie12}.  This
experiment used Ramsey's technique of separated oscillatory fields to
compare the precession rate of polarized cold neutrons in
a beam passing in close proximity to a polished copper plate with the
precession of neutrons in a reference beam.  The result in~\cite{pie12}, which
assumes no coupling to electrons ($f_{\perp}^{ne} = 0$) and
$f_{\perp}^{np} = f_{\perp}^{nn} \equiv f_{\perp}^{nN}$, is interpreted
as a limit on the coupling $(g_{A})^{2}$; the contribution from any
$g_{V}$ term is assumed negligible given the much stronger short--range
constraints on this parameter from torsion pendulum experiments with
unpolarized test masses.  The limit in Fig.~\ref{fig:up} ($\approx
(g_{A})^{2}/4$) has been re-scaled in accordance with these assumptions.

Similarly, the solid line in the $V_{12+13}$ plot is the limit on the
corresponding coupling for polarized neutrons derived from the neutron
spin rotation experiment at NIST~\cite{yan13}.  This experiment is
designed to be sensitive to the rotation $\phi$ of the polarization of a
transversely polarized beam of neutrons passing through a liquid
$^{4}$He target. The rotation $\phi$ arises from a $P$--violating
$\hat{\sigma}\cdot\hat{p}$ term in the forward scattering cross section, whether
induced by an interaction such as $V_{12+13}$ or the Standard Model weak
interaction to which the experiment is ultimately designed to be
sensitive.  The analysis in~\cite{yan13} uses the result on $\phi$,
currently an upper limit, to constrain $V_{12+13}$.  The limit is
reported in terms of $g_{V}g_{A}^{n}$, where $g_{V}$ contains a factor
$Z=2$ for $^{4}$He.  Equating the expression for $V_{12+13}$ in~\cite{yan13} 
to Eq.~\ref{eq:v1213} for polarized neutrons, 
and using $Z=2$, $A=4$ yields the result
($f_{v}^{ne}+f_{v}^{np}+f_{v}^{nn}=g_{V}g_{A}^{n}$) in
Fig.~\ref{fig:up}.  

The best limit on the $V_{9+10}$ interaction for electrons is 
derived from the Axion-Like Particle (ALP) torsion pendulum in the
Eot-Wash group, which consists of a thin silicon wafer
suspended between the two halves of a split toroidal
magnet~\cite{hoe11}.  The magnet provides the polarized electrons,
and the wafer a source of unpolarized nucleons highly insensitive to
the classical magnetic field present.  The limit in~\cite{hoe11}
is reported in terms of $g_{S}^{N}g_{P}^{e}$, where $g_{S}^{N}\equiv
g_{S}^{p}= g_{S}^{n} = g_{S}^{a}/A$ and it is assumed $g_{S}^{e} =
0$.  Since the unpolarized mass consists of silicon, the limits in
Fig.~\ref{fig:up} ($\feninet \approx 2g_{S}^{N}g_{P}^{e}$) are scaled according
to Table~\ref{tab:coup} with these assumptions, where the dashed line
is the projected thermal limit from~\cite{hoe11}.   The same scaling
applies to the prediction for the axion, shown in the $V_{9+10}$ plot
for the case of an $s=0$ interaction.  The prediction is again
from~\cite{moo84}, updated to account for the value of
$\theta_{QCD}$~\cite{cre79,*cre80} inferred from the current best limit on
the electric dipole moment of the neutron~\cite{bak06}. 

Finally, the $V_{9+10}$ plot in Fig.~\ref{fig:up} also shows the indirect
limits derived from a combination of data from laboratory experiments
and astrophysical arguments~\cite{raf12}.  These are limits on the coupling
$g_{S}^{N}g_{P}^{e}$, i.e., for the case of an $s=0$ interaction, in
which the constraints on $g_{S}^{N}$ come from short--range gravity
experiments with unpolarized test masses~\cite{kap07,ger08,sus11}, and
the limit on $g_{P}^{e}$ comes from stellar cooling.  They have been
scaled by the 
same factor in Fig.~\ref{fig:up} as the limit
in~\cite{hoe11} to maintain consistency with the results
in~\cite{raf12}.  As noted in~\cite{dob06}, analogous constraints on
$g_{S}^{N}g_{S}^{e}$ can be inferred by combining the same results for
$g_{S}^{N}$ with the stellar cooling limit on $g_{S}^{e}$.  Using
$g_{S}^{e} \leq 1.3\times 10^{-14}$ from~\cite{raf12}, the resulting limits
are shown in the $V_{4+5}$ plot for the case of an $s=0$ interaction.

\section{\label{sec:expt} Short--Range Experiment}
\begin{figure}
\includegraphics[width=3.3in]{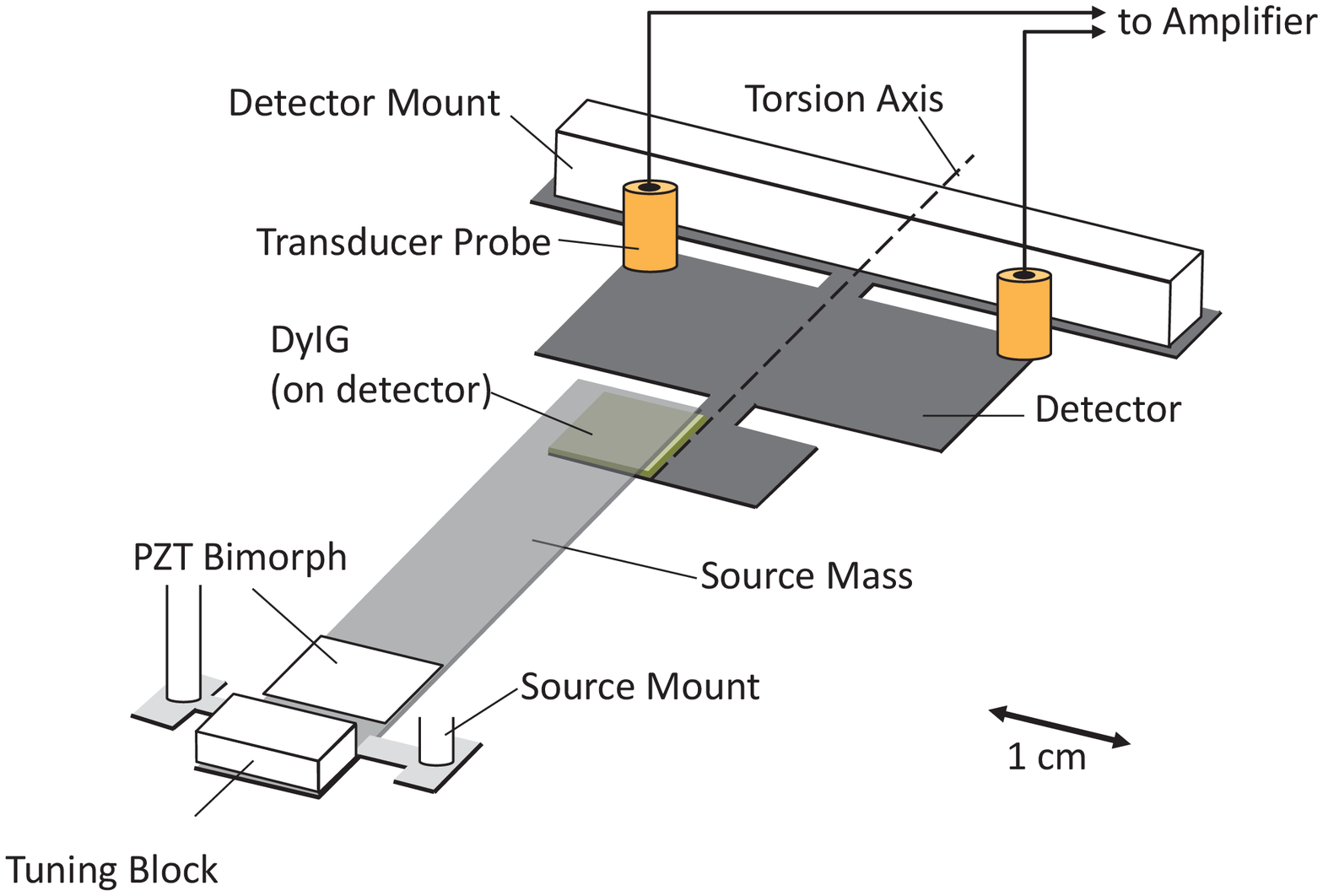}
  \caption{\label{fig:expt}(adapted from~\cite{lon03}
    and~\cite{yan14}) Basic test 
    mass geometry of the proposed
    experiment.  In the particular configuration, a thin sample of
    spin-polarized material (here DyIG) covers half of
    the small forward rectangle of the detector mass.  In 
    other configurations, a similar sample is attached to the
    underside of the forward part of the source mass.  The thin, stiff,
    conducting shield between the test masses is not shown.} 
\end{figure}

The experiment is illustrated in Fig.~\ref{fig:expt}.  It has
been used previously to set limits on mass--coupled
forces in the range of interest~\cite{lon03} and a more sensitive
version of it is currently fully operational.  The experimental test masses
consist of 1~kHz, planar mechanical oscillators with a thin shield between
them to suppress backgrounds.  The planar geometry is especially
efficient for concentrating as much mass as possible at the range of
interest.  It is nominally null with respect to $1/r^{2}$ forces and
thus effective in suppressing Newtonian backgrounds.  The (active)
source mass is driven at a resonance frequency of the (passive)
detector mass to maximize the signal.  For the mass--coupled force
search the test masses are made from tungsten, which has a density of
about 19~g/cm$^{3}$.  Resonant operation places a
heavy burden on vibration isolation.  The 1~kHz
operational frequency is chosen since in this frequency range it is possible
to construct a simple, passive vibration isolation system with high
dimensional stability~\cite{cha99}, permitting the test mass surfaces
to be maintained within a few microns of each other for indefinite
periods.

The source mass is a nodally--mounted cantilever driven by a
piezoelectric wafer attached in a region of high modal
curvature.  The detector is a planar double--torsional oscillator
originally developed for cryogenic 
condensed matter physics experiments~\cite{kle85,kli87}.  It consists
of 2 coplanar rectangles, joined along their central axes by a short
segment.  The resonant mode of interest is the first anti-symmetric
torsion mode, in which the rectangles
counter-rotate about the axis defined by the segment.  This mode is
distinguished by a high mechanical quality factor
($Q$), important for increasing sensitivity and suppressing thermal
noise.  To eliminate backgrounds mediated by electrostatic, residual gas, and 
possible Casimir effects, it is essential to place a stiff conducting
shield between the test masses.  The previous experiment~\cite{lon03}
used a 60 micron thick gold-coated sapphire plate clamped at two
opposite ends, which was completely effective at suppressing these
backgrounds.  The existing experiment uses a thinner shield made from
a stretched copper membrane.  Detector oscillations are read out with
a capacitive 
transducer coupled to a differential amplifier, which is sufficiently
sensitive to monitor the detector thermal motion~\cite{yan14}.

To make the experiment sensitive to spin--dependent
interactions, samples of spin--polarized materials can
be attached to the test masses (Fig.~\ref{fig:expt}).  The
principal challenges will be to fabricate such samples with the
necessary thin planar geometry while retaining the polarization, and
to control the extra backgrounds due to residual magnetic forces
that cannot be eliminated.  For the spin--polarized material,
compensated ferrimagnets are an intriguing possibility.  These
materials contain at least two magnetic sublattices in which the
magnetic moments are oppositely aligned.  The contributions of each
sub-lattice to the  magnetization of a sample depend on temperature in
such a way that there is a ``compensation'' temperature ($T_{c}$) at 
which their magnitudes are equal and thus cancel.  For materials in
which the contributions to the magnetism of each sub-lattice from spin
and orbital motion of the electrons are different, at the
compensation temperature there is a net spin.

The effect on the detector $Q$ of attaching a polarized sample is not
known.  However, silicon test mass prototypes, which are 
particularly attractive as low--susceptibility substrates for the
spin--dependent experiments, have been measured to have $Q$s as high
as $2\times 10^{6}$ between 77~K and room temperature.  For the purpose of the
sensitivity estimates, a conservative value of $Q=10000$ is
assumed. 

To locate the compensation temperature (assuming $T_{c} < 295$~K),
the experiment can be 
cooled radiatively with a high-emissivity shield surrounding the
central apparatus.  The test mass temperatures can be further adjusted
with thermoelectric elements.  The absolute magnetization
of the samples away from the compensation temperature, from which the
degree of spin--polarization can be deduced, can be measured using external
coils to produce a resonant, calibrated, quasi--uniform magnetic
gradient to drive the test masses.

\subsection*{\label{sec:garnet} Test mass development}
One candidate material for the polarized test masses,
Dy$_{6}$Fe$_{23}$, has been used in previous
experiments~\cite{rit90,ni93,chu93,ni94,hou03}. Dy$_{6}$Fe$_{23}$ is a
ferrimagnet with a net spin and a compensation temperature of about
250~K.  The Dy-Fe system exhibits several phases, however, and synthesis of the
pure 6-23 phase can be problematic~\cite{goo70,her84}. It
oxidizes readily and the samples in the reported
experiments are encapsulated, making it less attractive for fabrication of small
samples that must be kept in close proximity.  This work investigates
the rare earth iron garnets, in particular dysprosium iron garnet (DyIG),
Dy$_{3}^{3+}$Fe$_{2}^{3+}$Fe$_{3}^{3+}$O$_{12}$, as a possible
alternative.  The garnets are chemically stable and can be
produced in the lab with little difficulty. 

\subsubsection{Molecular field model}
DyIG is a ferrimagnet in which three sublattices contribute to the
magnetization.  The Dy$^{3+}$ ions occupy dodecahedral sites (commonly
denoted $c$) in the garnet lattice, the Fe$^{3+}$ octahedral sites
(denoted $a$) and tetrahedral sites (denoted $d$)~\cite{dio09}.  The
Dy$^{3+}$ moments are nominally aligned with the octahedral ion
moments and anti-aligned with the tetrahedral moments.  The total
magnetization per molecule $M$ at a particular temperature is thus: 
\begin{equation}
M = M_{c}+M_{a}-M_{d}.
\end{equation}

Following~\cite{dio09}, the contribution of each sublattice can be
calculated in a molecular field model. The temperature--dependent
sublattice moments are given by: 
\begin{eqnarray}
M_{c}(T) & = & M_{c}(0)B_{J_{c}}(x_{c})\nonumber \\
M_{a}(T) & = & M_{a}(0)B_{J_{a}}(x_{a})\nonumber \\
M_{d}(T) & = & M_{d}(0)B_{J_{d}}(x_{d}),
\label{eq:MT}
\end{eqnarray}    
where $M_{i}(0)$ are the 0~K moments and the $B_{J_{i}}(x_{i})$ are
the Brillouin functions for sublattice $i$.  For pure DyIG (that is,
no substitution of the ions on any sublattice) the 0~K moments are:  
\begin{eqnarray}
M_{c}(0) & = & 3g_{c}\mu_{B}J_{c} N_{A}\nonumber \\
M_{a}(0) & = & 2g_{a}\mu_{B}J_{a} N_{A}\nonumber \\
M_{d}(0) & = & 3g_{d}\mu_{B}J_{d} N_{A}.
\label{eq:M0}
\end{eqnarray}
Here, $\mu_{B}$ is the Bohr magneton in units of erg/Gauss and a factor of
Avogadro's number $N_{A}$ is included to convert $M$ to units of
$\mu_{B}$/molecule.  The coefficients in Eq.~\ref{eq:M0} represent the
relative numbers of $c$, $a$, and $d$ sites in the garnet
molecule~\cite{dio70}.  The terms $g_{i}$ and $J_{i}$ are the Lande
g-factor and total angular momentum of the ion on sublattice $i$.

The Boltzmann energy ratios $x_{i}$ in Eq.~\ref{eq:MT} are given by:
\begin{eqnarray}
x_{c} & = & \frac{g_{c}J_{c}\mu_{B}}{k_{B}T}[N_{cc}M_{c}+N_{ac}M_{a}+N_{cd}M_{d}]\nonumber \\
x_{a} & = & \frac{g_{a}J_{a}\mu_{B}}{k_{B}T}[N_{ac}M_{c}+N_{aa}M_{a}+N_{ad}M_{d}]\nonumber \\
x_{d} & = & \frac{g_{d}J_{d}\mu_{B}}{k_{B}T}[N_{cd}M_{c}+N_{ad}M_{a}+N_{dd}M_{d}],
\label{eq:x}
\end{eqnarray}
where the $N_{ij}$ are the molecular field coefficients.  Here, the exchange
fields (terms in brackets) are expressed in Gauss so that the $N_{ij}$
are in units of mol/cm$^{3}$.  With appropriate values of $g_{i}$ and
$J_{i}$, Eqs.~\ref{eq:MT} and~\ref{eq:x} are solved iteratively for
the three lattices
simultaneously.  The $N_{ij}$ are adjusted by trial and error to
reproduce the data on magnetization vs. temperature for pure DyIG
crystals.  

\begin{figure*}
\includegraphics[width=7.0in]{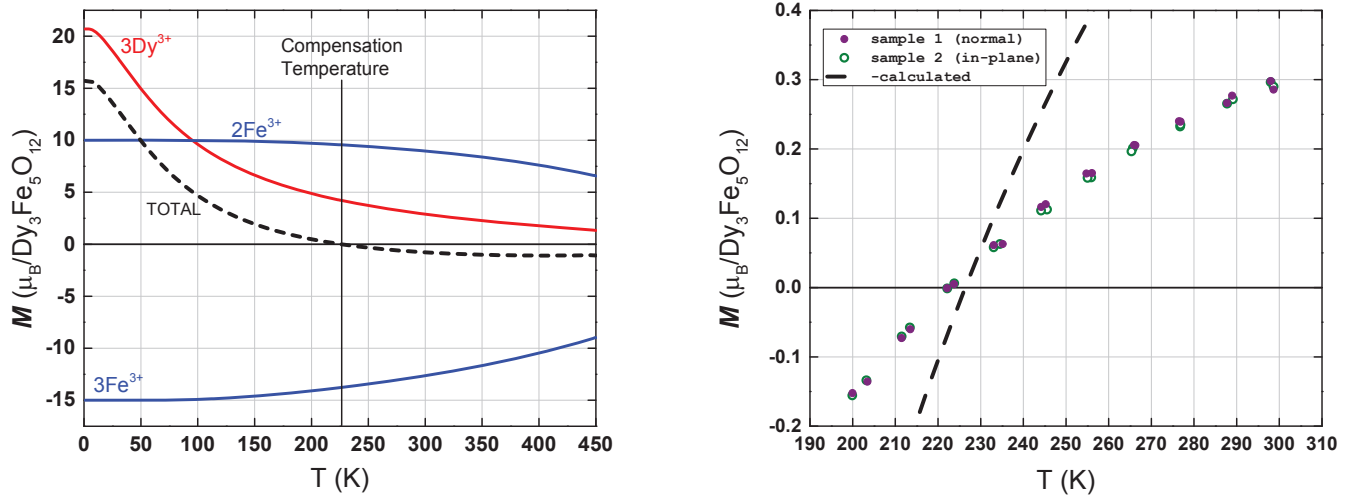}
  \caption{\label{fig:MT} \textit{Left:} Magnetization of
    DyIG versus
    temperature, from the results of the molecular field calculation
    described in the text.  Contributions from the ions on each
    sublattice are shown (bold curves), together with the sum (dashed
    curve). \textit{Right:} Magnetization vs temperature of the
    particular DyIG samples fabricated for the proposed
    experiment.  At each temperature investigated, there are
    two data points for each sample, one taken as the sample is
    cooled from 295~K, the other as the sample warmed from 200~K. The
    dashed line is the (negative of the) calculated total
    magnetization in the left-hand plot, for comparison with the slope
    of the experimental curves at $T{c}$.}   
\end{figure*}

Fig.~\ref{fig:MT} shows the result of the calculations of
magnetization vs. temperature using $N_{ac}=-4.0$~mol/cm$^3$,
$N_{dc}=6.0$~mol/cm$^3$~\cite{dio09}, $N_{cc}=0$~\cite{dio76}, and
$N_{aa}=-65.0$~mol/cm$^3$, $N_{ad}=97.0$~mol/cm$^3$, and
$N_{dd}=-30.4$~mol/cm$^3$~\cite{dio71}.  The calculation predicts
$T_{c}=226$~K.  At $T_{c}$, the three Dy$^{3+}$ ions contribute
4.1~$\mu_{B}$ to the magnetic moment of the DyIG molecule.  The two
Fe$^{3+}$ on the $a$ sublattice contribute 9.5~$\mu_{B}$, and the
three Fe$^{3+}$ on the $d$ sublattice contribute -13.6~$\mu_{B}$.  The
(absolute value of the) total magnetization curve displays good
agreement with data from measurements on single crystal spherical
samples~\cite{gel65}.

An analogous calculation for terbium iron garnet (TbIG), using the
appropriate $N_{ij}$ from the same references, predicts $T_{c}=266$~K.
The Tb$^{3+}$ ions contribute 4.0~$\mu_{B}$ to the total moment at $T_{c}$.

The Fe$^{3+}$ ions on the $a$ and $d$ sublattices have spin $S=5/2$ and
orbital angular momentum $L=0$.  Consequently, $J_{a}=J_{d}=5/2$ and
$g_{a}=g_{d}=2$ for the calculation in Fig.~\ref{fig:MT}.  It
should be noted that, for ions in the $3d^{n}$ series bonded in
an anion lattice such as garnet, the $3d$ shells are exposed to the
electrostatic fields of the lattice so that $L$ is uncoupled from $S$,
the process known as quenching.  A consequence is that $S$ is the
principal source of the magnetic moment and $g=2$ a good approximation
for most ions in this series.  The same effect has implications for the
correct values of $J_{c}$ and $g_{c}$.

The configuration of the Dy$^{3+}$ ion is $4f^{9}$. In contrast to the
Fe$^{3+}$ ions, the magnetically active $4f$ electrons in the rare earth are
shielded by the electrons in the full $5s$ and $5p$ outer
shells, thus they are not expected to be affected by the lattice
fields.  The free Dy$^{3+}$ ion has $S_{c}=5/2$ and $L_{c}=5$, for
$J_{c}=15/2$ and $g_{c}=4/3$.  However, these are not the values used
in the calculation in Fig.~\ref{fig:MT}.  To reproduce the data, the
effective value of $J_{c}$ is reduced, the process known as
canting.  Ref.~\cite{dio09} discusses two possible models.

In the first or semiclassical model, the $J_{c}$ vector is tilted with
respect to the direction defined by the spins of the $d$ lattice.  Thus
a projection $J_{c}^{\prime}=5.25$ is used in Eqs.~\ref{eq:MT}--\ref{eq:x},
together with $g_{c}=4/3$.  In the second model, $L_{c}$ is partially
quenched in the lattice field, leading to an actual reduction in
$J_{c}$.  Following the notation in~\cite{dio09}, the quenching factor
is $\gamma=0.38$, so that $L_{c}^{\prime\prime}=\gamma L_{c}=1.9$,
$J_{c}^{\prime\prime}=L_{c}^{\prime\prime}+S_{c}=4.4$, and
\[
g_{c}^{\prime\prime}=1+\frac{J_{c}^{\prime\prime}(J_{c}^{\prime\prime}+1)+S_{c}(S_{c}+1)-L_{c}^{\prime\prime}(L_{c}^{\prime\prime}+1)}{2J_{c}^{\prime\prime}(J_{c}^{\prime\prime}+1)}=1.57
\]
Either model produces the curves in Fig.~\ref{fig:MT}.\footnote{The
  values listed in~\cite{dio09} are $J_{c}^{\prime}=5.3$ and
  $\gamma=0.41$ (for $J_{c}^{\prime\prime}=4.6$ and
  $g_{c}^{\prime\prime}=1.54$). Use of these values in the
  authors' own calculation yields a prediction of $T_{c}=235$~K, in poorer
  agreement with the data in~\cite{gel65} and Fig.~\ref{fig:MT}.
  Presumably the differences can be accounted for by rounding in
  the calculations or of the
  reported values for $\gamma$, $J_{c}^{\prime}$, and the $N_{ij}$. Using
  either value of $\gamma$, the final results for the
  spin density of the samples are unchanged at the level of precision
  used.} However, as 
explained in~\cite{dio09}, the latter model with partially quenched
$L_{c}$ is more consistent with the results of measurements in fields applied
along the direction of the crystal fields.  It is also more
conservative for the purpose of estimating the spin excess of DyIG at
$T_{c}$, and thus is adopted here.

The spin contribution of the ions on the $i$th sublattice to the total
magnetic moment can be deduced from the spin g-factors, $g_{s_{i}}$. For the
Fe$^{3+}$ ions, which have $L=0$ and $g_{s}=g=2$, all of the contribution
is due to spin.  For the Dy$^{3+}$ ions in the lattice,
\[
g_{s}^{\prime\prime}=1+\frac{S_{c}(S_{c}+1)-L_{c}^{\prime\prime}(L_{c}^{\prime\prime}+1)}{J_{c}^{\prime\prime}(J_{c}^{\prime\prime}+1)}=1.14.
\]      
In this case, 73\% of the magnetic moment is due to spin and 27\% is
due to the orbital motion of the electrons. Thus, at $T_{c}$,
$\mu_{s_{c}}=3.1$~$\mu_{B}$ and the total spin excess per molecule (in
units of $\hbar$) is:
\begin{equation}
S_{T_{c}}=\frac{|\mu_{s_{T}}|}{2\mu_{B}}=\frac{|3.1+9.6-13.8|}{2}=0.6.
\end{equation}
The analogous calculation for TbIG ($L_{c}=S_{c}=3,
\gamma=0.32$~\cite{dio09}) yields $S_{T_{c}}=0.3$.  Thus while TbIG
may be more attractive for its higher $T_{c}$, the spin excess is
reduced by a factor of 2.

\subsubsection{Synthesis and properties}
Samples of DyIG practically sized for use in the proposed
experiment are synthesized via the chemical process described
in~\cite{ges94}.  The material is precipitated from a mixed metal
hydroxide precursor solution and dried in an oven (air atmosphere) at 393~K for 
12~hr.  It is then hand-ground to fine powder, and pressed (force =
10~kN) into 3.2~mm diameter pellets using a precision die mounted in a
hydraulic press.  The pellets are then fired in the oven at 1173~K for
18~hr.  Repetition of the grinding, pressing, and firing steps has been
shown to increase purity~\cite{ges94,uem08}; these steps were repeated
twice for the pellets in the present study.  Two such samples were fabricated,
sample 1 with thickness 0.84~mm and density 3.4~g/cm$^{3}$, sample 2
with thickness 0.97~mm and density 3.5~g/cm$^{3}$.

The sample magnetic properties were measured with a SQUID magnetometer
(Quantum Design MPMS--XL) calibrated with a palladium standard.  Both
samples were magnetized to saturation at room temperature in an applied
field of 2~T, then the applied field was ramped to zero.  Sample 1 was
magnetized in the direction normal to the plane of the pellet along
the symmetry axis, sample 2 was
magnetized in-plane (both polarizations are necessary for sensitivity
to all potentials in \vall, as explained in Sec.~\ref{sec:sens}).

The remnant magnetization of the samples was then measured as the
temperature was reduced below the anticipated $T_{c}$, then raised back
to room temperature. Results are shown in Fig.~\ref{fig:MT}. For both
samples, the magnetization drops to zero at a $T_{c}$ near 223~K,
reverses below, then recovers to the initial magnetization at room
temperature. Subsequent measurements show this behavior to be
repeatable upon multiple excursions through $T_{c}$, and when the
samples are held at $T_{c}$ for several hours.  Results are very
similar for the two polarizations, indicating little if any extra
demagnetization in the case of normal polarization.

The spin density of each sample at $T_{c}$ (assuming the density of
the pellets to be uniform) is given by:
\begin{equation}
n_{s}=\frac{N_{A}\rho}{A}S_{T_{c}},
\end{equation}
where $\rho$ is the mass density of the sample and $A=958.5$~g/mol is
the atomic weight of DyIG.  Following~\cite{rit90}, an additional
correction factor, equal to the ratio of the slope of the calculated
magnetization curve to the measured curves at $T_{c}$, is applied in order to
account for incomplete magnetization of the flat, polycrystalline
samples used. This ratio is 0.36, resulting in spin densities of
$n_{s}=4.0\times 10^{20}$~$\hbar$/cm$^{3}$ for sample 1 and $n_{s}=4.1\times
10^{20}$~$\hbar$/cm$^{3}$ for sample 2.  
%

\section{\label{sec:sens} Projected sensitivities}
The sensitivity of the experiment is based on the
expectation that essentially all experimental backgrounds can be
suppressed below the detector thermal noise and amplifier noise.  This
represents an ultimate 
practical sensitivity; results with reduced but competitive
sensitivity in the presence of other backgrounds are expected to be
realized sooner.

Experimental signals are estimated by converting
\vall\ to forces and integrating them
numerically over the test mass geometry, assuming values of 1
for the coupling constants.  For simplicity, it is assumed that each
of the interactions in \vall\ acts independently, as is the case for
the limits in Sec.~\ref{sec:limits}. (Additional limits on the
interactions in Eq.~\ref{eq:v11} are presented
in~\cite{hec13}, in which this assumption is relaxed.) 
The thermal noise force due to dissipation in the detector is found from
the mechanical Nyquist formula,
\begin{equation}
F_{T}=\sqrt\frac{4k_{B}Tm\omega_{0}}{Q\tau},
\end{equation}
where $k_{B}$ is Boltzmann's constant, $T$ is the temperature, $m$ is
the mass of the detector oscillator, $\omega_{0}$ is the resonance
frequency, $Q$ is the mechanical quality factor, and $\tau$ is the
experimental integration time. The ratio of this force to the result
of the integration of \vall\ at each
value of $\lambda$ used (that is, a signal--to--noise ratio of 1) yields
the sensitivity curves for the coupling constants. Since the
experiment is sensitive to changes in the signal as the test mass
separation is varied, the integration models the
sinusoidal modulation of the source mass and calculates the Fourier
amplitudes of the integrated signal.  In the thermal noise limit, the
amplitude of the oscillations of the detector is of order
$\sqrt{k_{B}T/(m \omega_{0}^{2})} \sim 1$~pm (Table~\ref{tab:geo}),
thus the relative velocity term $\vec{v}$ in \vall\ is very well
approximated by the source velocity.

To maximize sensitivity at short range, a small but reasonable minimum
test mass gap (that is, distance of closest approach) is assumed.  This is fixed
at 120~$\mu$m. This allows for a 100~$\mu$m
thick shield between the test masses (40~$\mu$m thicker than the
shield used successfully in previous experiments~\cite{lon03}, thus
reserving space for additional magnetic shielding if needed).  For
each value of $\lambda$ investigated, the source mass amplitude is
optimized for maximum signal.  For the static interactions in \vall,
this results in values of order $\lambda$.  For all $\lambda$ above
1~mm, an amplitude of 1~mm is used, which is taken to represent a
practical maximum with the piezoelectric drive technique.  The
optimization is the same for the velocity-dependent
interactions.  The exceptions are $V_{8}$ and $V_{16}$,
which, on account of the $v^{2}$ dependence, increase monotonically
with source amplitude at any $\lambda$ over the range of interest.  For these
interactions, the maximum practical source amplitude of 1~mm is used at
each value of $\lambda$.\footnote{On account of the $v^{2}$
  dependence, the principal signals for $V_{8}$ and $V_{16}$ are at
  twice the source frequency, for source amplitudes below $\lambda$.  Given the
narrow detector resonance at $\omega_{0}$, sensitivity 
to these interactions is maximized by driving the source at
$\omega_{0}/2$.  Since the corresponding reduction in source velocity
leads to a reduction of the signal by a factor of 4, this is practical
only for the case when the second harmonic exceeds the fundamental by
more than a factor of 4, which is true only for $\lambda > 0.5$~mm.  Generally,
since the sinusoidal source velocity scales with amplitude,
higher harmonics of all other velocity--dependent potentials exceed
the size of the optimized fundamental as the source amplitude
increases.  The excess is never more than a factor of 2, however, and
cannot be exploited for additional sensitivity.}

\begin{figure}
\includegraphics[width=3.3in]{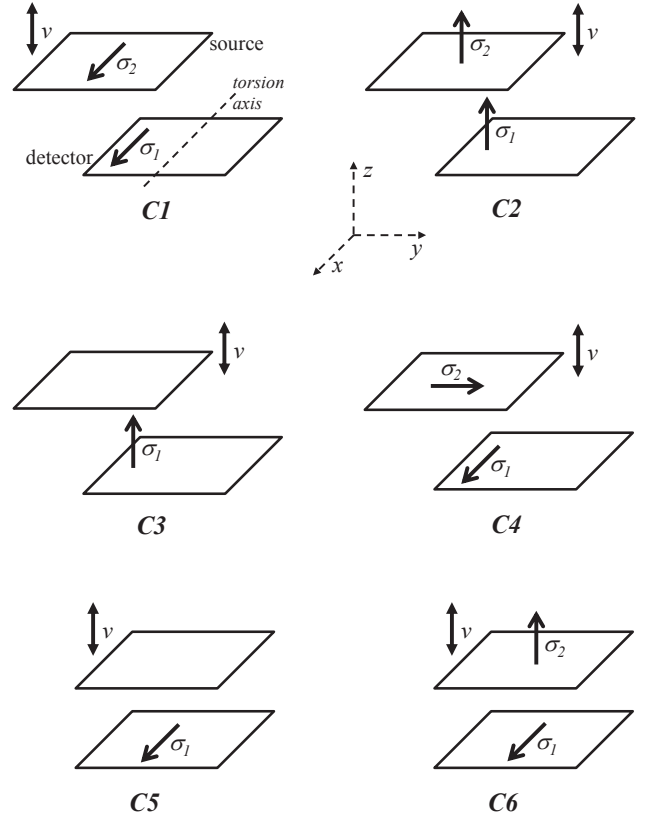}
  \caption{\label{fig:configs} Test mass and spin polarization
    configurations used to
    search for interactions $V_{2}$--$V_{16}$ as assumed in the
    sensitivity calculations. Here, $\sigma_{1}$ is the net polarization
    direction of the spins in the detector mass and $\sigma_{2}$ in the source
    mass.  The relative velocity of the spins in each test mass is
    strongly dominated by the velocity of the source, $v$. The
    detector torsion axis is along $x$. $C1$:
    polarization in--plane, parallel ($V_{2}$, $V_{3}$). $C2$:
    polarization normal ($V_{2}$, $V_{3}$, $V_{6+7}$, $V_{8}$). $C3$:
    polarization normal, detector only ($V_{9+10}$, $V_{12+13}$). $C4$:
    polarization in--plane, crossed ($V_{11}$, $V_{14}$). $C5$:
    polarization in--plane, detector only ($V_{4+5}$).  $C6$:
    polarization mixed ($V_{15}$, $V_{16}$). Note that in
    $C_{1}$--$C_{4}$, the source subtends half the detector area and
    the polarized material $\sigma_{1}$ covers only the detector area
    subtended. In $C_{5}$ and $C_{6}$, the source mass is centered over
    the detector and $\sigma_{1}$ covers the entire detector area.} 
\end{figure}

Sensitivity to all interactions in \vall\ is possible in principle with
simple modifications to the test mass geometry and polarization.  The
different configurations are illustrated in
Fig.~\ref{fig:configs}.  For the purposes of the sensitivity
calculations, pure vertical translation of the source mass (along the
$z$-axis in Fig.~\ref{fig:configs}) is assumed, with instantaneous
velocity $v$. This is 
a good approximation for the planar geometry but there will be small
corrections for the actual mode
shape of a practical source mass. The six
configurations include four in which the spin--polarized material
covers only half of the detector mass and the source mass is
positioned over that half, so that the resulting force is optimized to
excite the sensitive torsional mode of the detector: 
\begin{itemize}
\item[C1:] Detector and source polarization in--plane and
  parallel. Presumably the 
  easiest configuration to attain for spin--spin interactions and
  sensitive to potentials proportional to $\sdots$. 
\item[C2:] Polarization normal to the test mass
  planes and parallel to $\vec{v}$, for optimum sensitivity to $\sdots$ and
  spin--spin interactions proportional to $\sdotr$ and $\sdotv$.   
\item[C3:] Polarization normal (detector only) and parallel to $\vec{v}$,
  for optimum sensitivity to 
  spin--mass interactions proportional to $\sdotr$ and $\sdotv$.
\item[C4:] Polarization in--plane and crossed, for sensitivity to
  spin--spin interactions proportional to $(\scrosss) \cdot \hat{r}$ and
  $(\scrosss) \cdot \vec{v}$.
\end{itemize}
In the two remaining configurations, the polarized material covers the
entire detector surface and the source is centered over the detector,
for sensitivity to interactions proportional to $\vcrossr$.  The
$\vcrossr$ term averages to zero over the 
surface of the detector in this configuration, however, the
associated vector field has the profile of a vortex centered in the
detector plane. Thus, for $\hat{\sigma}_{1}$ parallel to the detector
torsion axis, a force proportional to $\hat{\sigma}_{1}\cdot
(\vcrossr)$, while averaging to zero over the entire detector plane,
averages to a non-zero value on one side of the torsion axis and the
negative of this value on the other, efficiently driving the
torsional mode of interest:
\begin{itemize}     
\item[C5:] Polarization in--plane, parallel to detector torsion axis,
  for sensitivity to spin--mass interactions proportional to
  $\hat{\sigma}\cdot (\vcrossr)$.  
\item[C6:] Polarization mixed, with one parallel to detector torsion axis,
  for sensitivity to spin--spin interactions proportional to
  $[\hat{\sigma}_{1,2}\cdot (\vcrossr)] (\hat{\sigma}_{2,1}\cdot
  \hat{r})$ and  
  $[\hat{\sigma}_{1,2}\cdot (\vcrossr)] (\hat{\sigma}_{2,1}\cdot \vec{v})$.  
\end{itemize}
Parameters used in the sensitivity calculations are listed in
Table.~\ref{tab:geo}.
\begin{table}[htb]
\caption{\label{tab:geo} Test mass geometry and other properties used
  in sensitivity 
  calculations. For searches in which the source is centered over the
  detector ($V_{4+5}, V_{15}, V_{16}$), the active detector area is
  58~mm$^{2}$. For interactions $V_{4+5}, V_{9+10}$, and $V_{12+13}$,
  the source is unpolarized and consists of silicon (density
  2.3~g/cm$^{3}$).}   
\begin{ruledtabular} 
\begin{tabular}{lc}
\multicolumn{1}{c}{\textbf{Parameter}}&\multicolumn{1}{c}{\textbf{value}}\\ \hline
Active detector area          &29~mm$^{2}$ \\
Active source mass area       &36~mm$^{2}$ \\ 
Test mass thickness           &1~mm \\ 
Test mass density             &3.5~g/cm$^{3}$ (DyIG) \\ 
Test mass spin density        &$4 \times 10^{20}/$cm$^{3}$ \\   
Minimum source-detector gap   &120~$\mu$m \\
Signal frequency              &1~kHz\\            
Detector quality factor       &$1 \times 10^{4}$ \\ 
Temperature                   &225~K \\
Integration time              &200~hr\\                     
\end{tabular}
\end{ruledtabular}
\end{table}

Results for sensitivity to the static spin--spin interactions
(Eq.~\ref{eq:v11}) are shown in Fig.~\ref{fig:ppstatic}. 
The sensitivity to the $V_{2}$ interaction is comparable to
the Eot--Wash and UVA experiments in the range near 1~cm, but many
orders more so only a few millimeters below on the account of the
small test mass separation.  

The projected limit on $V_{2}$ is the most sensitive relative to the
others, by at least 4 orders of magnitude, in the range of interest.
The remaining projections can be roughly grouped into three regions of
successively decreasing sensitivity, determined by the number of
additional factors of $1/r$ or $v/c$ in the expressions for the corresponding
interactions (Eqs.~\ref{eq:v11}-\ref{eq:v1213}) relative to $V_{2}$.

The sensitivity to the $V_{3}$ dipole-dipole
interaction is about eight orders of magnitude
greater than the limit inferred from positronium spectroscopy at
20~$\mu$m. Results for sensitivity to
the velocity--dependent spin--spin interactions (Eq.~\ref{eq:v16}) are
shown in Fig.~\ref{fig:ppdyn}.  The proposed technique would appear to
have unique sensitivity in this range.

Results for sensitivity to interactions between polarized electrons
and unpolarized atoms (Eq.~\ref{eq:v1213}) are shown in
Fig.~\ref{fig:up}.  The sensitivity to the $V_{9+10}$
monopole--dipole interaction is about eight orders of magnitude
greater than the current experimental limits at 20~$\mu$m.
The lower dashed curves in the $V_{4+5}$ and $V_{9+10}$ plots are the
projected limit on $g_{S}^{N}g_{S}^{e}$ and $g_{S}^{N}g_{P}^{e}$, respectively, 
using the value for $g_{S}^{e}$ and $g_{P}^{e}$ from stellar
cooling~\cite{raf12} and the projected limit on $g_{S}^{N}$ from the version of
the proposed experiment using dense, unpolarized test
masses~\cite{lon03b}.

\begin{acknowledgments}
The authors would like to thank H.-O. Meyer for contributions to the
analysis of spin--polarized materials, D. Sprinke and R. Manus for
assistance with the magnetization measurements, and B. Dobrescu, W.-T. Ni,
and W. M. Snow for useful discussions.  This work is supported by
National Science Foundation grant PHY-1207656, and the Indiana
University Center for Spacetime Symmetries (IUCSS). T. L. acknowledges
the support of the Indiana University Cox Scholarship Program.
\end{acknowledgments}

\bibliography{long_spin_sens_201406r}

\providecommand{\noopsort}[1]{}\providecommand{\singleletter}[1]{#1}%
\begin{thebibliography}{52}%
\makeatletter
\providecommand \@ifxundefined [1]{%
 \@ifx{#1\undefined}
}%
\providecommand \@ifnum [1]{%
 \ifnum #1\expandafter \@firstoftwo
 \else \expandafter \@secondoftwo
 \fi
}%
\providecommand \@ifx [1]{%
 \ifx #1\expandafter \@firstoftwo
 \else \expandafter \@secondoftwo
 \fi
}%
\providecommand \natexlab [1]{#1}%
\providecommand \enquote  [1]{``#1''}%
\providecommand \bibnamefont  [1]{#1}%
\providecommand \bibfnamefont [1]{#1}%
\providecommand \citenamefont [1]{#1}%
\providecommand \href@noop [0]{\@secondoftwo}%
\providecommand \href [0]{\begingroup \@sanitize@url \@href}%
\providecommand \@href[1]{\@@startlink{#1}\@@href}%
\providecommand \@@href[1]{\endgroup#1\@@endlink}%
\providecommand \@sanitize@url [0]{\catcode `\\12\catcode `\$12\catcode
  `\&12\catcode `\#12\catcode `\^12\catcode `\_12\catcode `\%12\relax}%
\providecommand \@@startlink[1]{}%
\providecommand \@@endlink[0]{}%
\providecommand \url  [0]{\begingroup\@sanitize@url \@url }%
\providecommand \@url [1]{\endgroup\@href {#1}{\urlprefix }}%
\providecommand \urlprefix  [0]{URL }%
\providecommand \Eprint [0]{\href }%
\providecommand \doibase [0]{http://dx.doi.org/}%
\providecommand \selectlanguage [0]{\@gobble}%
\providecommand \bibinfo  [0]{\@secondoftwo}%
\providecommand \bibfield  [0]{\@secondoftwo}%
\providecommand \translation [1]{[#1]}%
\providecommand \BibitemOpen [0]{}%
\providecommand \bibitemStop [0]{}%
\providecommand \bibitemNoStop [0]{.\EOS\space}%
\providecommand \EOS [0]{\spacefactor3000\relax}%
\providecommand \BibitemShut  [1]{\csname bibitem#1\endcsname}%
\let\auto@bib@innerbib\@empty
\bibitem [{\citenamefont {Beringer}\ \emph {et~al.}(2012)\citenamefont
  {Beringer} \emph {et~al.}}]{ber12}%
  \BibitemOpen
  \bibfield  {author} {\bibinfo {author} {\bibfnamefont {J.}~\bibnamefont
  {Beringer}} \emph {et~al.} (\bibinfo {collaboration} {Particle Data Group}),\
  }\href@noop {} {\bibfield  {journal} {\bibinfo  {journal} {Phys.\ Rev. D}\
  }\textbf {\bibinfo {volume} {86}},\ \bibinfo {pages} {010001 and 2013 partial
  update for the 2014 edition} (\bibinfo {year} {2012})}\BibitemShut {NoStop}%
\bibitem [{\citenamefont {Adelberger}\ \emph {et~al.}(2009)\citenamefont
  {Adelberger}, \citenamefont {Gundlach}, \citenamefont {Heckel}, \citenamefont
  {Hoedl},\ and\ \citenamefont {Schlamminger}}]{ade09}%
  \BibitemOpen
  \bibfield  {author} {\bibinfo {author} {\bibfnamefont {E.~G.}\ \bibnamefont
  {Adelberger}}, \bibinfo {author} {\bibfnamefont {J.~H.}\ \bibnamefont
  {Gundlach}}, \bibinfo {author} {\bibfnamefont {B.~R.}\ \bibnamefont
  {Heckel}}, \bibinfo {author} {\bibfnamefont {S.}~\bibnamefont {Hoedl}}, \
  and\ \bibinfo {author} {\bibfnamefont {S.}~\bibnamefont {Schlamminger}},\
  }\href@noop {} {\bibfield  {journal} {\bibinfo  {journal} {Prog. Part. Nucl.
  Phys.}\ }\textbf {\bibinfo {volume} {62}},\ \bibinfo {pages} {102} (\bibinfo
  {year} {2009})}\BibitemShut {NoStop}%
\bibitem [{\citenamefont {Jaeckel}\ and\ \citenamefont
  {Ringwald}(2010)}]{jae10}%
  \BibitemOpen
  \bibfield  {author} {\bibinfo {author} {\bibfnamefont {J.}~\bibnamefont
  {Jaeckel}}\ and\ \bibinfo {author} {\bibfnamefont {A.}~\bibnamefont
  {Ringwald}},\ }\href@noop {} {\bibfield  {journal} {\bibinfo  {journal} {Ann.
  Rev. Nucl. Part. Sci.}\ }\textbf {\bibinfo {volume} {60}},\ \bibinfo {pages}
  {405} (\bibinfo {year} {2010})}\BibitemShut {NoStop}%
\bibitem [{\citenamefont {Pospelov}\ \emph {et~al.}(2008)\citenamefont
  {Pospelov}, \citenamefont {Ritz},\ and\ \citenamefont {Voloshin}}]{pos08}%
  \BibitemOpen
  \bibfield  {author} {\bibinfo {author} {\bibfnamefont {M.}~\bibnamefont
  {Pospelov}}, \bibinfo {author} {\bibfnamefont {A.}~\bibnamefont {Ritz}}, \
  and\ \bibinfo {author} {\bibfnamefont {M.~B.}\ \bibnamefont {Voloshin}},\
  }\href@noop {} {\bibfield  {journal} {\bibinfo  {journal} {Phys. Lett. B}\
  }\textbf {\bibinfo {volume} {662}},\ \bibinfo {pages} {53} (\bibinfo {year}
  {2008})}\BibitemShut {NoStop}%
\bibitem [{\citenamefont {Dobrescu}\ and\ \citenamefont
  {Mocioiu}(2006)}]{dob06}%
  \BibitemOpen
  \bibfield  {author} {\bibinfo {author} {\bibfnamefont {B.}~\bibnamefont
  {Dobrescu}}\ and\ \bibinfo {author} {\bibfnamefont {I.}~\bibnamefont
  {Mocioiu}},\ }\href@noop {} {\bibfield  {journal} {\bibinfo  {journal} {J.
  High Energy Phys.}\ }\textbf {\bibinfo {volume} {11}},\ \bibinfo {pages}
  {005} (\bibinfo {year} {2006})}\BibitemShut {NoStop}%
\bibitem [{\citenamefont {Hunter}\ and\ \citenamefont {Ang}(2013)}]{hun13}%
  \BibitemOpen
  \bibfield  {author} {\bibinfo {author} {\bibfnamefont {L.~R.}\ \bibnamefont
  {Hunter}}\ and\ \bibinfo {author} {\bibfnamefont {D.~G.}\ \bibnamefont
  {Ang}},\ }\href@noop {} {\bibfield  {journal} {\bibinfo  {journal} {Phys.
  Rev. Lett.}\ }\textbf {\bibinfo {volume} {112}},\ \bibinfo {pages} {091803}
  (\bibinfo {year} {2013})}\BibitemShut {NoStop}%
\bibitem [{\citenamefont {Kimball}\ \emph {et~al.}(2010)\citenamefont
  {Kimball}, \citenamefont {Boyd},\ and\ \citenamefont {Budker}}]{jac10}%
  \BibitemOpen
  \bibfield  {author} {\bibinfo {author} {\bibfnamefont {D.~F.~J.}\
  \bibnamefont {Kimball}}, \bibinfo {author} {\bibfnamefont {A.}~\bibnamefont
  {Boyd}}, \ and\ \bibinfo {author} {\bibfnamefont {D.}~\bibnamefont
  {Budker}},\ }\href@noop {} {\bibfield  {journal} {\bibinfo  {journal} {Phys.
  Rev. A}\ }\textbf {\bibinfo {volume} {82}},\ \bibinfo {pages} {062714}
  (\bibinfo {year} {2010})}\BibitemShut {NoStop}%
\bibitem [{\citenamefont {Long}\ \emph {et~al.}(2003)\citenamefont {Long},
  \citenamefont {Chan}, \citenamefont {Churnside}, \citenamefont {Gulbis},
  \citenamefont {Varney},\ and\ \citenamefont {Price}}]{lon03}%
  \BibitemOpen
  \bibfield  {author} {\bibinfo {author} {\bibfnamefont {J.~C.}\ \bibnamefont
  {Long}}, \bibinfo {author} {\bibfnamefont {H.~W.}\ \bibnamefont {Chan}},
  \bibinfo {author} {\bibfnamefont {A.~B.}\ \bibnamefont {Churnside}}, \bibinfo
  {author} {\bibfnamefont {E.~A.}\ \bibnamefont {Gulbis}}, \bibinfo {author}
  {\bibfnamefont {M.~C.~M.}\ \bibnamefont {Varney}}, \ and\ \bibinfo {author}
  {\bibfnamefont {J.~C.}\ \bibnamefont {Price}},\ }\href@noop {} {\bibfield
  {journal} {\bibinfo  {journal} {Nature}\ }\textbf {\bibinfo {volume} {421}},\
  \bibinfo {pages} {922} (\bibinfo {year} {2003})}\BibitemShut {NoStop}%
\bibitem [{\citenamefont {Moody}\ and\ \citenamefont {Wilczek}(1984)}]{moo84}%
  \BibitemOpen
  \bibfield  {author} {\bibinfo {author} {\bibfnamefont {J.~E.}\ \bibnamefont
  {Moody}}\ and\ \bibinfo {author} {\bibfnamefont {F.}~\bibnamefont
  {Wilczek}},\ }\href@noop {} {\bibfield  {journal} {\bibinfo  {journal} {Phys.
  Rev. D}\ }\textbf {\bibinfo {volume} {30}},\ \bibinfo {pages} {130} (\bibinfo
  {year} {1984})}\BibitemShut {NoStop}%
\bibitem [{\citenamefont {Long}\ and\ \citenamefont {Price}(2003)}]{lon03b}%
  \BibitemOpen
  \bibfield  {author} {\bibinfo {author} {\bibfnamefont {J.~C.}\ \bibnamefont
  {Long}}\ and\ \bibinfo {author} {\bibfnamefont {J.~C.}\ \bibnamefont
  {Price}},\ }\href@noop {} {\bibfield  {journal} {\bibinfo  {journal} {C. R.
  Physique}\ }\textbf {\bibinfo {volume} {4}},\ \bibinfo {pages} {337}
  (\bibinfo {year} {2003})}\BibitemShut {NoStop}%
\bibitem [{\citenamefont {Heckel}\ \emph {et~al.}(2008)\citenamefont {Heckel},
  \citenamefont {Adelberger}, \citenamefont {Cramer}, \citenamefont {Cook},
  \citenamefont {Schlamminger},\ and\ \citenamefont {Schmidt}}]{hec08}%
  \BibitemOpen
  \bibfield  {author} {\bibinfo {author} {\bibfnamefont {B.~R.}\ \bibnamefont
  {Heckel}}, \bibinfo {author} {\bibfnamefont {E.~G.}\ \bibnamefont
  {Adelberger}}, \bibinfo {author} {\bibfnamefont {C.~E.}\ \bibnamefont
  {Cramer}}, \bibinfo {author} {\bibfnamefont {T.~S.}\ \bibnamefont {Cook}},
  \bibinfo {author} {\bibfnamefont {S.}~\bibnamefont {Schlamminger}}, \ and\
  \bibinfo {author} {\bibfnamefont {U.}~\bibnamefont {Schmidt}},\ }\href@noop
  {} {\bibfield  {journal} {\bibinfo  {journal} {Phys. Rev. D}\ }\textbf
  {\bibinfo {volume} {78}},\ \bibinfo {pages} {092006} (\bibinfo {year}
  {2008})}\BibitemShut {NoStop}%
\bibitem [{\citenamefont {Heckel}\ \emph {et~al.}(2013)\citenamefont {Heckel},
  \citenamefont {Terrano},\ and\ \citenamefont {Adelberger}}]{hec13}%
  \BibitemOpen
  \bibfield  {author} {\bibinfo {author} {\bibfnamefont {B.~R.}\ \bibnamefont
  {Heckel}}, \bibinfo {author} {\bibfnamefont {W.~A.}\ \bibnamefont {Terrano}},
  \ and\ \bibinfo {author} {\bibfnamefont {E.~G.}\ \bibnamefont {Adelberger}},\
  }\href@noop {} {\bibfield  {journal} {\bibinfo  {journal} {Phys. Rev. Lett.}\
  }\textbf {\bibinfo {volume} {111}},\ \bibinfo {pages} {151802} (\bibinfo
  {year} {2013})}\BibitemShut {NoStop}%
\bibitem [{\citenamefont {Ritter}\ \emph {et~al.}(1990)\citenamefont {Ritter},
  \citenamefont {Goldblum}, \citenamefont {Ni}, \citenamefont {Gillies},\ and\
  \citenamefont {Speake}}]{rit90}%
  \BibitemOpen
  \bibfield  {author} {\bibinfo {author} {\bibfnamefont {R.~C.}\ \bibnamefont
  {Ritter}}, \bibinfo {author} {\bibfnamefont {C.~E.}\ \bibnamefont
  {Goldblum}}, \bibinfo {author} {\bibfnamefont {W.-T.}\ \bibnamefont {Ni}},
  \bibinfo {author} {\bibfnamefont {G.~T.}\ \bibnamefont {Gillies}}, \ and\
  \bibinfo {author} {\bibfnamefont {C.~C.}\ \bibnamefont {Speake}},\
  }\href@noop {} {\bibfield  {journal} {\bibinfo  {journal} {Phys. Rev. D}\
  }\textbf {\bibinfo {volume} {42}},\ \bibinfo {pages} {977} (\bibinfo {year}
  {1990})}\BibitemShut {NoStop}%
\bibitem [{\citenamefont {Ni}\ \emph {et~al.}(1994)\citenamefont {Ni},
  \citenamefont {Chui}, \citenamefont {Pan},\ and\ \citenamefont
  {Cheng}}]{ni94}%
  \BibitemOpen
  \bibfield  {author} {\bibinfo {author} {\bibfnamefont {W.-T.}\ \bibnamefont
  {Ni}}, \bibinfo {author} {\bibfnamefont {T.~C.~P.}\ \bibnamefont {Chui}},
  \bibinfo {author} {\bibfnamefont {S.-S.}\ \bibnamefont {Pan}}, \ and\
  \bibinfo {author} {\bibfnamefont {B.-Y.}\ \bibnamefont {Cheng}},\ }\href@noop
  {} {\bibfield  {journal} {\bibinfo  {journal} {Physica B}\ }\textbf {\bibinfo
  {volume} {194}},\ \bibinfo {pages} {153} (\bibinfo {year}
  {1994})}\BibitemShut {NoStop}%
\bibitem [{\citenamefont {Mills}\ and\ \citenamefont {Bearman}(1975)}]{mil75}%
  \BibitemOpen
  \bibfield  {author} {\bibinfo {author} {\bibfnamefont {A.~P.}\ \bibnamefont
  {Mills}}\ and\ \bibinfo {author} {\bibfnamefont {G.~H.}\ \bibnamefont
  {Bearman}},\ }\href@noop {} {\bibfield  {journal} {\bibinfo  {journal} {Phys.
  Rev. Lett.}\ }\textbf {\bibinfo {volume} {34}},\ \bibinfo {pages} {246}
  (\bibinfo {year} {1975})}\BibitemShut {NoStop}%
\bibitem [{\citenamefont {Hill}(2001)}]{hil01}%
  \BibitemOpen
  \bibfield  {author} {\bibinfo {author} {\bibfnamefont {R.~J.}\ \bibnamefont
  {Hill}},\ }\href@noop {} {\bibfield  {journal} {\bibinfo  {journal} {Phys.
  Rev. Lett.}\ }\textbf {\bibinfo {volume} {86}},\ \bibinfo {pages} {3280}
  (\bibinfo {year} {2001})}\BibitemShut {NoStop}%
\bibitem [{\citenamefont {Dobrescu}()}]{dob13}%
  \BibitemOpen
  \bibfield  {author} {\bibinfo {author} {\bibfnamefont {B.}~\bibnamefont
  {Dobrescu}},\ }\href@noop {} {}\bibinfo {howpublished} {private
  communication}\BibitemShut {NoStop}%
\bibitem [{\citenamefont {Ni}\ \emph {et~al.}(1993)\citenamefont {Ni},
  \citenamefont {Pan}, \citenamefont {Chui},\ and\ \citenamefont
  {Cheng}}]{ni93}%
  \BibitemOpen
  \bibfield  {author} {\bibinfo {author} {\bibfnamefont {W.-T.}\ \bibnamefont
  {Ni}}, \bibinfo {author} {\bibfnamefont {S.-S.}\ \bibnamefont {Pan}},
  \bibinfo {author} {\bibfnamefont {T.~C.~P.}\ \bibnamefont {Chui}}, \ and\
  \bibinfo {author} {\bibfnamefont {B.-Y.}\ \bibnamefont {Cheng}},\ }\href@noop
  {} {\bibfield  {journal} {\bibinfo  {journal} {Int. J. Mod. Phys. A}\
  }\textbf {\bibinfo {volume} {8}},\ \bibinfo {pages} {5153} (\bibinfo {year}
  {1993})}\BibitemShut {NoStop}%
\bibitem [{\citenamefont {Chui}\ and\ \citenamefont {Ni}(1993)}]{chu93}%
  \BibitemOpen
  \bibfield  {author} {\bibinfo {author} {\bibfnamefont {T.~C.~P.}\
  \bibnamefont {Chui}}\ and\ \bibinfo {author} {\bibfnamefont {W.-T.}\
  \bibnamefont {Ni}},\ }\href@noop {} {\bibfield  {journal} {\bibinfo
  {journal} {Phys. Rev. Lett.}\ }\textbf {\bibinfo {volume} {71}},\ \bibinfo
  {pages} {3247} (\bibinfo {year} {1993})}\BibitemShut {NoStop}%
\bibitem [{\citenamefont {Ni}()}]{ni13}%
  \BibitemOpen
  \bibfield  {author} {\bibinfo {author} {\bibfnamefont {W.-T.}\ \bibnamefont
  {Ni}},\ }\href@noop {} {}\bibinfo {howpublished} {private
  communication}\BibitemShut {NoStop}%
\bibitem [{\citenamefont {Mills}(1983)}]{mil83}%
  \BibitemOpen
  \bibfield  {author} {\bibinfo {author} {\bibfnamefont {A.~P.}\ \bibnamefont
  {Mills}},\ }\href@noop {} {\bibfield  {journal} {\bibinfo  {journal} {Phys.
  Rev. A}\ }\textbf {\bibinfo {volume} {27}},\ \bibinfo {pages} {262} (\bibinfo
  {year} {1983})}\BibitemShut {NoStop}%
\bibitem [{\citenamefont {Ritter}\ \emph {et~al.}(1984)\citenamefont {Ritter},
  \citenamefont {Egan}, \citenamefont {Hughes},\ and\ \citenamefont
  {Woodle}}]{rit84}%
  \BibitemOpen
  \bibfield  {author} {\bibinfo {author} {\bibfnamefont {M.~W.}\ \bibnamefont
  {Ritter}}, \bibinfo {author} {\bibfnamefont {P.~O.}\ \bibnamefont {Egan}},
  \bibinfo {author} {\bibfnamefont {V.~W.}\ \bibnamefont {Hughes}}, \ and\
  \bibinfo {author} {\bibfnamefont {K.~A.}\ \bibnamefont {Woodle}},\
  }\href@noop {} {\bibfield  {journal} {\bibinfo  {journal} {Phys. Rev. A}\
  }\textbf {\bibinfo {volume} {30}},\ \bibinfo {pages} {1331} (\bibinfo {year}
  {1984})}\BibitemShut {NoStop}%
\bibitem [{\citenamefont {Kniehl}\ and\ \citenamefont {Penin}(2000)}]{kni00}%
  \BibitemOpen
  \bibfield  {author} {\bibinfo {author} {\bibfnamefont {B.~A.}\ \bibnamefont
  {Kniehl}}\ and\ \bibinfo {author} {\bibfnamefont {A.~A.}\ \bibnamefont
  {Penin}},\ }\href@noop {} {\bibfield  {journal} {\bibinfo  {journal} {Phys.
  Rev. Lett.}\ }\textbf {\bibinfo {volume} {85}},\ \bibinfo {pages} {5094}
  (\bibinfo {year} {2000})}\BibitemShut {NoStop}%
\bibitem [{\citenamefont {Melnikov}\ and\ \citenamefont
  {Yelkhovsky}(2001)}]{mel01}%
  \BibitemOpen
  \bibfield  {author} {\bibinfo {author} {\bibfnamefont {K.}~\bibnamefont
  {Melnikov}}\ and\ \bibinfo {author} {\bibfnamefont {A.}~\bibnamefont
  {Yelkhovsky}},\ }\href@noop {} {\bibfield  {journal} {\bibinfo  {journal}
  {Phys. Rev. Lett.}\ }\textbf {\bibinfo {volume} {86}},\ \bibinfo {pages}
  {1498} (\bibinfo {year} {2001})}\BibitemShut {NoStop}%
\bibitem [{\citenamefont {Liao}\ and\ \citenamefont {Liu}(2007)}]{lia07}%
  \BibitemOpen
  \bibfield  {author} {\bibinfo {author} {\bibfnamefont {Y.}~\bibnamefont
  {Liao}}\ and\ \bibinfo {author} {\bibfnamefont {J.-Y.}\ \bibnamefont {Liu}},\
  }\href@noop {} {\bibfield  {journal} {\bibinfo  {journal} {Phys. Rev. Lett.}\
  }\textbf {\bibinfo {volume} {99}},\ \bibinfo {pages} {191804} (\bibinfo
  {year} {2007})}\BibitemShut {NoStop}%
\bibitem [{\citenamefont {Rosenberg}\ and\ \citenamefont {van
  Bibber}(2000)}]{ros00}%
  \BibitemOpen
  \bibfield  {author} {\bibinfo {author} {\bibfnamefont {L.~J.}\ \bibnamefont
  {Rosenberg}}\ and\ \bibinfo {author} {\bibfnamefont {K.~A.}\ \bibnamefont
  {van Bibber}},\ }\href@noop {} {\bibfield  {journal} {\bibinfo  {journal}
  {Phys. Rep.}\ }\textbf {\bibinfo {volume} {325}},\ \bibinfo {pages} {1}
  (\bibinfo {year} {2000})}\BibitemShut {NoStop}%
\bibitem [{\citenamefont {Visinelli}\ and\ \citenamefont {Gondolo}()}]{vis14}%
  \BibitemOpen
  \bibfield  {author} {\bibinfo {author} {\bibfnamefont {L.}~\bibnamefont
  {Visinelli}}\ and\ \bibinfo {author} {\bibfnamefont {P.}~\bibnamefont
  {Gondolo}},\ }\href@noop {} {}\Eprint
  {http://arxiv.org/abs/arXiv:1403.4594v2} {arXiv:1403.4594v2} \BibitemShut
  {NoStop}%
\bibitem [{\citenamefont {Ade}\ \emph {et~al.}()\citenamefont {Ade} \emph
  {et~al.}}]{bic14}%
  \BibitemOpen
  \bibfield  {author} {\bibinfo {author} {\bibfnamefont {P.~A.~R.}\
  \bibnamefont {Ade}} \emph {et~al.} (\bibinfo {collaboration} {BICEP2
  Collaboration}),\ }\href@noop {} {}\Eprint
  {http://arxiv.org/abs/arXiv:1403.3985} {arXiv:1403.3985} \BibitemShut
  {NoStop}%
\bibitem [{\citenamefont {Piegsa}\ and\ \citenamefont {Pignol}(2012)}]{pie12}%
  \BibitemOpen
  \bibfield  {author} {\bibinfo {author} {\bibfnamefont {F.~M.}\ \bibnamefont
  {Piegsa}}\ and\ \bibinfo {author} {\bibfnamefont {G.}~\bibnamefont
  {Pignol}},\ }\href@noop {} {\bibfield  {journal} {\bibinfo  {journal} {Phys.
  Rev. Lett.}\ }\textbf {\bibinfo {volume} {108}},\ \bibinfo {pages} {181801}
  (\bibinfo {year} {2012})}\BibitemShut {NoStop}%
\bibitem [{\citenamefont {Yan}\ and\ \citenamefont {Snow}(2013)}]{yan13}%
  \BibitemOpen
  \bibfield  {author} {\bibinfo {author} {\bibfnamefont {H.}~\bibnamefont
  {Yan}}\ and\ \bibinfo {author} {\bibfnamefont {W.~M.}\ \bibnamefont {Snow}},\
  }\href@noop {} {\bibfield  {journal} {\bibinfo  {journal} {Phys. Rev. Lett.}\
  }\textbf {\bibinfo {volume} {110}},\ \bibinfo {pages} {082003} (\bibinfo
  {year} {2013})}\BibitemShut {NoStop}%
\bibitem [{\citenamefont {Hoedl}\ \emph {et~al.}(2011)\citenamefont {Hoedl},
  \citenamefont {Fleischer}, \citenamefont {Adelberger},\ and\ \citenamefont
  {Heckel}}]{hoe11}%
  \BibitemOpen
  \bibfield  {author} {\bibinfo {author} {\bibfnamefont {S.~A.}\ \bibnamefont
  {Hoedl}}, \bibinfo {author} {\bibfnamefont {F.}~\bibnamefont {Fleischer}},
  \bibinfo {author} {\bibfnamefont {E.~G.}\ \bibnamefont {Adelberger}}, \ and\
  \bibinfo {author} {\bibfnamefont {B.~R.}\ \bibnamefont {Heckel}},\
  }\href@noop {} {\bibfield  {journal} {\bibinfo  {journal} {Phys. Rev. Lett.}\
  }\textbf {\bibinfo {volume} {106}},\ \bibinfo {pages} {041801} (\bibinfo
  {year} {2011})}\BibitemShut {NoStop}%
\bibitem [{\citenamefont {Raffelt}(2012)}]{raf12}%
  \BibitemOpen
  \bibfield  {author} {\bibinfo {author} {\bibfnamefont {G.}~\bibnamefont
  {Raffelt}},\ }\href@noop {} {\bibfield  {journal} {\bibinfo  {journal} {Phys.
  Rev. D}\ }\textbf {\bibinfo {volume} {86}},\ \bibinfo {pages} {015001}
  (\bibinfo {year} {2012})}\BibitemShut {NoStop}%
\bibitem [{\citenamefont {Crewther}\ \emph {et~al.}(1979)\citenamefont
  {Crewther}, \citenamefont {Vecchia}, \citenamefont {Veneziano},\ and\
  \citenamefont {Witten}}]{cre79}%
  \BibitemOpen
  \bibfield  {author} {\bibinfo {author} {\bibfnamefont {R.~J.}\ \bibnamefont
  {Crewther}}, \bibinfo {author} {\bibfnamefont {P.~D.}\ \bibnamefont
  {Vecchia}}, \bibinfo {author} {\bibfnamefont {G.}~\bibnamefont {Veneziano}},
  \ and\ \bibinfo {author} {\bibfnamefont {E.}~\bibnamefont {Witten}},\
  }\href@noop {} {\bibfield  {journal} {\bibinfo  {journal} {Phys. Lett. B}\
  }\textbf {\bibinfo {volume} {88}},\ \bibinfo {pages} {123} (\bibinfo {year}
  {1979})}\BibitemShut {NoStop}%
\bibitem [{\citenamefont {Crewther}\ \emph {et~al.}(1980)\citenamefont
  {Crewther}, \citenamefont {Vecchia}, \citenamefont {Veneziano},\ and\
  \citenamefont {Witten}}]{cre80}%
  \BibitemOpen
  \bibfield  {author} {\bibinfo {author} {\bibfnamefont {R.~J.}\ \bibnamefont
  {Crewther}}, \bibinfo {author} {\bibfnamefont {P.~D.}\ \bibnamefont
  {Vecchia}}, \bibinfo {author} {\bibfnamefont {G.}~\bibnamefont {Veneziano}},
  \ and\ \bibinfo {author} {\bibfnamefont {E.}~\bibnamefont {Witten}},\
  }\href@noop {} {\bibfield  {journal} {\bibinfo  {journal} {Phys. Lett. B}\
  }\textbf {\bibinfo {volume} {91}},\ \bibinfo {pages} {487} (\bibinfo {year}
  {1980})}\BibitemShut {NoStop}%
\bibitem [{\citenamefont {Baker}\ \emph {et~al.}(2006)\citenamefont {Baker}
  \emph {et~al.}}]{bak06}%
  \BibitemOpen
  \bibfield  {author} {\bibinfo {author} {\bibfnamefont {C.~A.}\ \bibnamefont
  {Baker}} \emph {et~al.},\ }\href@noop {} {\bibfield  {journal} {\bibinfo
  {journal} {Phys. Rev. Lett.}\ }\textbf {\bibinfo {volume} {97}},\ \bibinfo
  {pages} {131801} (\bibinfo {year} {2006})}\BibitemShut {NoStop}%
\bibitem [{\citenamefont {Kapner}\ \emph {et~al.}(2007)\citenamefont {Kapner},
  \citenamefont {Cook}, \citenamefont {Adelberger}, \citenamefont {Gundlach},
  \citenamefont {Heckel}, \citenamefont {Hoyle},\ and\ \citenamefont
  {Swanson}}]{kap07}%
  \BibitemOpen
  \bibfield  {author} {\bibinfo {author} {\bibfnamefont {D.~J.}\ \bibnamefont
  {Kapner}}, \bibinfo {author} {\bibfnamefont {T.~S.}\ \bibnamefont {Cook}},
  \bibinfo {author} {\bibfnamefont {E.~G.}\ \bibnamefont {Adelberger}},
  \bibinfo {author} {\bibfnamefont {J.~H.}\ \bibnamefont {Gundlach}}, \bibinfo
  {author} {\bibfnamefont {B.~R.}\ \bibnamefont {Heckel}}, \bibinfo {author}
  {\bibfnamefont {C.~D.}\ \bibnamefont {Hoyle}}, \ and\ \bibinfo {author}
  {\bibfnamefont {H.~E.}\ \bibnamefont {Swanson}},\ }\href@noop {} {\bibfield
  {journal} {\bibinfo  {journal} {Phys. Rev. Lett}\ }\textbf {\bibinfo {volume}
  {98}},\ \bibinfo {pages} {021101} (\bibinfo {year} {2007})}\BibitemShut
  {NoStop}%
\bibitem [{\citenamefont {Geraci}\ \emph {et~al.}(2008)\citenamefont {Geraci},
  \citenamefont {Smullin}, \citenamefont {Weld}, \citenamefont {Chiaverini},\
  and\ \citenamefont {Kapitulnik}}]{ger08}%
  \BibitemOpen
  \bibfield  {author} {\bibinfo {author} {\bibfnamefont {A.~A.}\ \bibnamefont
  {Geraci}}, \bibinfo {author} {\bibfnamefont {S.~J.}\ \bibnamefont {Smullin}},
  \bibinfo {author} {\bibfnamefont {D.~M.}\ \bibnamefont {Weld}}, \bibinfo
  {author} {\bibfnamefont {J.}~\bibnamefont {Chiaverini}}, \ and\ \bibinfo
  {author} {\bibfnamefont {A.}~\bibnamefont {Kapitulnik}},\ }\href@noop {}
  {\bibfield  {journal} {\bibinfo  {journal} {Phys. Rev. D}\ }\textbf {\bibinfo
  {volume} {78}},\ \bibinfo {pages} {022002} (\bibinfo {year}
  {2008})}\BibitemShut {NoStop}%
\bibitem [{\citenamefont {Sushkov}\ \emph {et~al.}(2011)\citenamefont
  {Sushkov}, \citenamefont {Kim}, \citenamefont {Dalvit},\ and\ \citenamefont
  {Lamoreaux}}]{sus11}%
  \BibitemOpen
  \bibfield  {author} {\bibinfo {author} {\bibfnamefont {A.~O.}\ \bibnamefont
  {Sushkov}}, \bibinfo {author} {\bibfnamefont {W.~J.}\ \bibnamefont {Kim}},
  \bibinfo {author} {\bibfnamefont {D.~A.~R.}\ \bibnamefont {Dalvit}}, \ and\
  \bibinfo {author} {\bibfnamefont {S.~K.}\ \bibnamefont {Lamoreaux}},\
  }\href@noop {} {\bibfield  {journal} {\bibinfo  {journal} {Phys. Rev. Lett.}\
  }\textbf {\bibinfo {volume} {107}},\ \bibinfo {pages} {171101} (\bibinfo
  {year} {2011})}\BibitemShut {NoStop}%
\bibitem [{\citenamefont {Chan}\ \emph {et~al.}(1999)\citenamefont {Chan},
  \citenamefont {Long},\ and\ \citenamefont {Price}}]{cha99}%
  \BibitemOpen
  \bibfield  {author} {\bibinfo {author} {\bibfnamefont {H.~W.}\ \bibnamefont
  {Chan}}, \bibinfo {author} {\bibfnamefont {J.~C.}\ \bibnamefont {Long}}, \
  and\ \bibinfo {author} {\bibfnamefont {J.~C.}\ \bibnamefont {Price}},\
  }\href@noop {} {\bibfield  {journal} {\bibinfo  {journal} {Rev. Sci.
  Instrum.}\ }\textbf {\bibinfo {volume} {70}},\ \bibinfo {pages} {2742}
  (\bibinfo {year} {1999})}\BibitemShut {NoStop}%
\bibitem [{\citenamefont {Yan}\ \emph {et~al.}(2014)\citenamefont {Yan},
  \citenamefont {Housworth}, \citenamefont {Meyer}, \citenamefont {Visser},
  \citenamefont {Weisman},\ and\ \citenamefont {Long}}]{yan14}%
  \BibitemOpen
  \bibfield  {author} {\bibinfo {author} {\bibfnamefont {H.}~\bibnamefont
  {Yan}}, \bibinfo {author} {\bibfnamefont {E.}~\bibnamefont {Housworth}},
  \bibinfo {author} {\bibfnamefont {H.-O.}\ \bibnamefont {Meyer}}, \bibinfo
  {author} {\bibfnamefont {G.}~\bibnamefont {Visser}}, \bibinfo {author}
  {\bibfnamefont {E.}~\bibnamefont {Weisman}}, \ and\ \bibinfo {author}
  {\bibfnamefont {J.~C.}\ \bibnamefont {Long}},\ }\href@noop {} {} (\bibinfo
  {year} {2014}),\ \bibinfo {note} {{Phys. Rev. D} (submitted)},\ \Eprint
  {http://arxiv.org/abs/arXiv:1402.0145} {arXiv:1402.0145} \BibitemShut
  {NoStop}%
\bibitem [{\citenamefont {Kleiman}\ \emph {et~al.}(1985)\citenamefont
  {Kleiman}, \citenamefont {Kaminsky}, \citenamefont {Reppy}, \citenamefont
  {Pindak},\ and\ \citenamefont {Bishop}}]{kle85}%
  \BibitemOpen
  \bibfield  {author} {\bibinfo {author} {\bibfnamefont {R.~N.}\ \bibnamefont
  {Kleiman}}, \bibinfo {author} {\bibfnamefont {G.~K.}\ \bibnamefont
  {Kaminsky}}, \bibinfo {author} {\bibfnamefont {J.~D.}\ \bibnamefont {Reppy}},
  \bibinfo {author} {\bibfnamefont {R.}~\bibnamefont {Pindak}}, \ and\ \bibinfo
  {author} {\bibfnamefont {D.~J.}\ \bibnamefont {Bishop}},\ }\href@noop {}
  {\bibfield  {journal} {\bibinfo  {journal} {Rev. Sci. Instrum.}\ }\textbf
  {\bibinfo {volume} {56}},\ \bibinfo {pages} {2088} (\bibinfo {year}
  {1985})}\BibitemShut {NoStop}%
\bibitem [{\citenamefont {Klitsner}\ and\ \citenamefont {Pohl}(1986)}]{kli87}%
  \BibitemOpen
  \bibfield  {author} {\bibinfo {author} {\bibfnamefont {T.}~\bibnamefont
  {Klitsner}}\ and\ \bibinfo {author} {\bibfnamefont {R.~O.}\ \bibnamefont
  {Pohl}},\ }\href@noop {} {\bibfield  {journal} {\bibinfo  {journal} {Phys.
  Rev. B}\ }\textbf {\bibinfo {volume} {34}},\ \bibinfo {pages} {6045}
  (\bibinfo {year} {1986})}\BibitemShut {NoStop}%
\bibitem [{\citenamefont {Hou}\ \emph {et~al.}(2003)\citenamefont {Hou},
  \citenamefont {Ni},\ and\ \citenamefont {Li}}]{hou03}%
  \BibitemOpen
  \bibfield  {author} {\bibinfo {author} {\bibfnamefont {L.-S.}\ \bibnamefont
  {Hou}}, \bibinfo {author} {\bibfnamefont {W.-T.}\ \bibnamefont {Ni}}, \ and\
  \bibinfo {author} {\bibfnamefont {Y.-C.~M.}\ \bibnamefont {Li}},\ }\href@noop
  {} {\bibfield  {journal} {\bibinfo  {journal} {Phys. Rev. Lett.}\ }\textbf
  {\bibinfo {volume} {90}},\ \bibinfo {pages} {201101} (\bibinfo {year}
  {2003})}\BibitemShut {NoStop}%
\bibitem [{\citenamefont {van~der Goot}\ and\ \citenamefont
  {Buschow}(1970)}]{goo70}%
  \BibitemOpen
  \bibfield  {author} {\bibinfo {author} {\bibfnamefont {A.~S.}\ \bibnamefont
  {van~der Goot}}\ and\ \bibinfo {author} {\bibfnamefont {K.~H.~J.}\
  \bibnamefont {Buschow}},\ }\href@noop {} {\bibfield  {journal} {\bibinfo
  {journal} {J. Less-Common Metals}\ }\textbf {\bibinfo {volume} {21}},\
  \bibinfo {pages} {151} (\bibinfo {year} {1970})}\BibitemShut {NoStop}%
\bibitem [{\citenamefont {Herbst}\ and\ \citenamefont {Croat}(1984)}]{her84}%
  \BibitemOpen
  \bibfield  {author} {\bibinfo {author} {\bibfnamefont {J.}~\bibnamefont
  {Herbst}}\ and\ \bibinfo {author} {\bibfnamefont {J.}~\bibnamefont {Croat}},\
  }\href@noop {} {\bibfield  {journal} {\bibinfo  {journal} {J. Appl. Phys.}\
  }\textbf {\bibinfo {volume} {55}},\ \bibinfo {pages} {3023} (\bibinfo {year}
  {1984})}\BibitemShut {NoStop}%
\bibitem [{\citenamefont {Dionne}(2009)}]{dio09}%
  \BibitemOpen
  \bibfield  {author} {\bibinfo {author} {\bibfnamefont {G.~F.}\ \bibnamefont
  {Dionne}},\ }\href@noop {} {\emph {\bibinfo {title} {Magnetic Oxides}}}\
  (\bibinfo  {publisher} {Springer},\ \bibinfo {address} {New York},\ \bibinfo
  {year} {2009})\BibitemShut {NoStop}%
\bibitem [{\citenamefont {Dionne}(1970)}]{dio70}%
  \BibitemOpen
  \bibfield  {author} {\bibinfo {author} {\bibfnamefont {G.~F.}\ \bibnamefont
  {Dionne}},\ }\href@noop {} {\bibfield  {journal} {\bibinfo  {journal} {J.
  Appl. Phys.}\ }\textbf {\bibinfo {volume} {41}},\ \bibinfo {pages} {4874}
  (\bibinfo {year} {1970})}\BibitemShut {NoStop}%
\bibitem [{\citenamefont {Dionne}(1976)}]{dio76}%
  \BibitemOpen
  \bibfield  {author} {\bibinfo {author} {\bibfnamefont {G.~F.}\ \bibnamefont
  {Dionne}},\ }\href@noop {} {\bibfield  {journal} {\bibinfo  {journal} {J.
  Appl. Phys.}\ }\textbf {\bibinfo {volume} {47}},\ \bibinfo {pages} {4220}
  (\bibinfo {year} {1976})}\BibitemShut {NoStop}%
\bibitem [{\citenamefont {Dionne}(1971)}]{dio71}%
  \BibitemOpen
  \bibfield  {author} {\bibinfo {author} {\bibfnamefont {G.~F.}\ \bibnamefont
  {Dionne}},\ }\href@noop {} {\bibfield  {journal} {\bibinfo  {journal} {J.
  Appl. Phys.}\ }\textbf {\bibinfo {volume} {42}},\ \bibinfo {pages} {2142}
  (\bibinfo {year} {1971})}\BibitemShut {NoStop}%
\bibitem [{\citenamefont {Geller}\ \emph {et~al.}(1965)\citenamefont {Geller},
  \citenamefont {Remeika}, \citenamefont {Sherwood}, \citenamefont {Williams},\
  and\ \citenamefont {Espinosa}}]{gel65}%
  \BibitemOpen
  \bibfield  {author} {\bibinfo {author} {\bibfnamefont {S.}~\bibnamefont
  {Geller}}, \bibinfo {author} {\bibfnamefont {J.~P.}\ \bibnamefont {Remeika}},
  \bibinfo {author} {\bibfnamefont {R.~C.}\ \bibnamefont {Sherwood}}, \bibinfo
  {author} {\bibfnamefont {H.~J.}\ \bibnamefont {Williams}}, \ and\ \bibinfo
  {author} {\bibfnamefont {G.~P.}\ \bibnamefont {Espinosa}},\ }\href@noop {}
  {\bibfield  {journal} {\bibinfo  {journal} {Phys. Rev.}\ }\textbf {\bibinfo
  {volume} {137}},\ \bibinfo {pages} {A1034} (\bibinfo {year}
  {1965})}\BibitemShut {NoStop}%
\bibitem [{\citenamefont {Gesselbracht}\ \emph {et~al.}(1994)\citenamefont
  {Gesselbracht}, \citenamefont {Cappellari}, \citenamefont {Ellis},
  \citenamefont {Rzeznik},\ and\ \citenamefont {Johnson}}]{ges94}%
  \BibitemOpen
  \bibfield  {author} {\bibinfo {author} {\bibfnamefont {M.~J.}\ \bibnamefont
  {Gesselbracht}}, \bibinfo {author} {\bibfnamefont {A.~M.}\ \bibnamefont
  {Cappellari}}, \bibinfo {author} {\bibfnamefont {A.~B.}\ \bibnamefont
  {Ellis}}, \bibinfo {author} {\bibfnamefont {M.~R.}\ \bibnamefont {Rzeznik}},
  \ and\ \bibinfo {author} {\bibfnamefont {B.~R.}\ \bibnamefont {Johnson}},\
  }\href@noop {} {\bibfield  {journal} {\bibinfo  {journal} {J. Chem. Educ.}\
  }\textbf {\bibinfo {volume} {71}},\ \bibinfo {pages} {696} (\bibinfo {year}
  {1994})}\BibitemShut {NoStop}%
\bibitem [{\citenamefont {Uemura}\ \emph {et~al.}(2008)\citenamefont {Uemura},
  \citenamefont {Yamagishi}, \citenamefont {Ebisu}, \citenamefont {Chikazawa},\
  and\ \citenamefont {Nagata}}]{uem08}%
  \BibitemOpen
  \bibfield  {author} {\bibinfo {author} {\bibfnamefont {M.}~\bibnamefont
  {Uemura}}, \bibinfo {author} {\bibfnamefont {T.}~\bibnamefont {Yamagishi}},
  \bibinfo {author} {\bibfnamefont {S.}~\bibnamefont {Ebisu}}, \bibinfo
  {author} {\bibfnamefont {S.}~\bibnamefont {Chikazawa}}, \ and\ \bibinfo
  {author} {\bibfnamefont {S.}~\bibnamefont {Nagata}},\ }\href@noop {}
  {\bibfield  {journal} {\bibinfo  {journal} {Phil. Mag.}\ }\textbf {\bibinfo
  {volume} {88}},\ \bibinfo {pages} {209} (\bibinfo {year} {2008})}\BibitemShut
  {NoStop}%
\end{thebibliography}%

\end{document}